\documentclass[12pt]{article}

\usepackage{epsfig} 
\usepackage{epstopdf}
\usepackage{amssymb}
\usepackage{amsmath}
\usepackage{amsfonts}
\usepackage{graphicx}
\graphicspath{ {plots/} }
\usepackage{mathrsfs}
\usepackage{color}
\usepackage{multirow}

\newcommand{\C}{\mathbb{C}}

\newcommand{\fa}{\mathfrak{a}}
\newcommand{\fb}{\mathfrak{b}}

\newcommand{\fn}{{\mathfrak{n}}}

\newcommand{\fK}{\mathfrak{K}}

\newcommand{\bba}{\mathbf{a}}
\newcommand{\bb}{\mathbf{b}}

\newcommand{\bt}{\mathbf{t}}

\newcommand{\bz}{\mathbf{z}}

\newcommand{\be}{\begin{equation}}
\newcommand{\ee}{\end{equation}}
\newcommand{\bea}{\begin{eqnarray}}
\newcommand{\eea}{\end{eqnarray}}
\newcommand{\nn}{\nonumber}

\newcommand{\ed}{\end{document}}

\newcommand{\bi}{\begin{itemize}}
\newcommand{\ei}{\end{itemize}}

\newcommand{\bce}{\begin{center}}
\newcommand{\ece}{\end{center}}

\newcommand{\sE}{\mathscr{E}}

\newcommand{\sR}{\mathscr{R}}

\newcommand{\sT}{\mathscr{T}}

\newcommand{\RE}{{\rm Re}}
\newcommand{\IM}{{\rm Im}}

\begin{document}

\title{Spectral Singularities, Threshold Gain, and Output Intensity for a Slab Laser with Mirrors}

\author{Keremcan Do\u{g}an$^1$, Ali~Mostafazadeh$^{1,2,}$\thanks{Corresponding author, Email Address: amostafazadeh@ku.edu.tr} , and Mustafa Sar{\i}saman$^1$\\[6pt]
Departments of Physics$^1$ and Mathematics$^2$, Ko\c{c} University,\\ 34450 Sar{\i}yer,
Istanbul, Turkey}

\date{ }
\maketitle

\begin{abstract}

We explore the consequences of the emergence of linear and nonlinear spectral singularities in TE modes of a homogeneous slab of active optical material that is placed between two mirrors. We use the results to derive explicit expressions for the laser threshold condition and laser output intensity for these modes of the slab and discuss their physical implications. In particular, we reveal the details of the dependence of the threshold gain and output intensity on the position and properties of the mirrors and on the real part of the refractive index of the gain material.

\vspace{2mm}



\end{abstract}

\section{Introduction}
\label{Sec1}

\begin{figure}[t]
    \begin{center}
    \includegraphics[scale=.55]{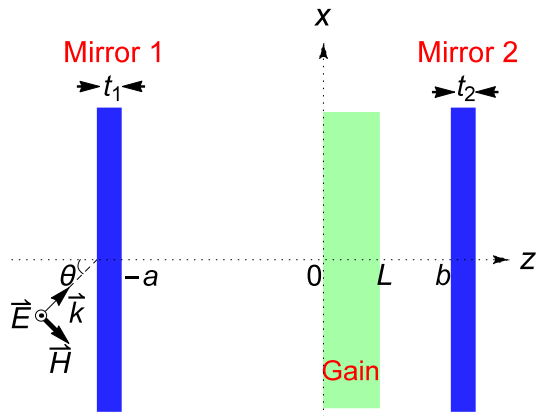}
    \caption{(Color online) Schematic representation of a planar slab of thickness $L$ that is made out of a homogeneous gain material, and surrounded by mirrors.}
    \label{fig1}
    \end{center}
    \end{figure}

A slab of homogeneous gain material that is placed between a pair of parallel mirrors, as depicted in Fig.~\ref{fig1},
provides a simple model for a laser. The mirrors act as the boundaries of a Fabry Perot resonator that enhance the optical path of the wave inside the gain region, hence reducing the threshold gain. The laser oscillations start once the gain coefficient $g$ for the system exceeds its threshold value $g_0$. The produced laser light can then exit the system through the mirror with higher transmission (lower reflection) coefficient.

In principle regardless of the details and geometry of a laser, in our case the type of the gain material and the presence and properties of the mirrors, laser light emission involves the generation of outgoing coherent electromagnetic waves. Purely outgoing solutions of a wave equation define its resonances  \cite{Siegert}. If such a solution is required not to decay and behave like a scattering solution, i.e., tend to a plane wave in spatial infinities, then its wavenumber and consequently its energy must be real. Recalling that the imaginary part of the energy of a resonance determines its width, we can relate laser light emission to certain zero-width resonances. Ref.~\cite{prl-2009} identifies these with the mathematical notion of a spectral singularity \cite{naimark,kemp,schwartz}. A remarkable outcome of this identification is that the existence of a spectral singularity for an optical potential describing an active system coincides with the laser threshold condition for the system,  i.e., $g=g_0$ for a homogenous slab laser. An explicit demonstration of this result is provided in Ref.~\cite{pra-2011a} for normally incident TE modes of a mirrorless slab laser and subsequently used as a computational scheme for the determination of the threshold gain for bilayer gain media \cite{bilayer1,bilayer2} and media with spherical or cylindrical geometries that lase in their radial \cite{spherical} or whispering gallery modes \cite{WGM}. 

Another interesting application of spectral singularities is in the study of coherent perfect absorption (antilasing) \cite{antilaser1,antilaser2,longhi-CPA,longhi-2010}. A system serves as a coherent perfect absorber provided that the complex-conjugate of the associated optical potential has a spectral singularity \cite{bilayer1}; in effect spectral singularities provide the basic mathematical tool for describing lasers and antilasers. For further discussion of physical aspects and applications of spectral singularities, see \cite{aalipour,Longhi,CP-2012,GC-2013-14,HR,Li-2014,pra-2015a,wang-2016,Hang-2016,Kalozoumis-2016,Pendharker-2016,jiang-2016}.

Another characteristic feature of lasing is its nonlinear nature \cite{lamb}. Once the gain coefficient exceeds its threshold value and the laser oscillations begin, the propagation of purely outgoing waves inside the gain medium leads to a nonlinear response of the medium. This observation has motivated a generalization of the notion of spectral singularity to nonlinear wave equations \cite{prl-2013,liu,reddy}. Just above the threshold the nonlinearity can be treated as a first-order perturbation of the relevant Helmholtz equation. Refs.~\cite{pra-2013c,sam-2014,p131} show that if we identify this perturbation with a weak Kerr nonlinearity, then the condition for the emergence of a nonlinear spectral singularity yields an expression for the laser output intensity $I$ that is linear in the gain coefficient $g$. More specifically,
    \be
    I=\left(\frac{g-g_0}{\sigma g_0}\right)\widehat I,
    \label{eq1}
    \ee
where $\sigma$ is the Kerr coefficient, and $\widehat I$ is a function of the geometry and other parameters of the system. The linear dependence of $I$ on $g$ is one of the basic results of laser physics. Here it follows from the purely mathematical condition of the existence of a nonlinear spectral singularity \cite{prl-2013}.

In the present article, we explore the linear and nonlinear spectral singularities in the TE modes of a homogenous slab of gain material that is placed between a pair of parallel mirrors. Our aim is to determine the explicit form of the threshold gain $g_0$ and the slope $\widehat I$ of the intensity without having to rely on any physical arguments. More precisely, we offer a derivation of the laser threshold condition and the linear-dependence of the intensity on the gain coefficient using the following two basic postulates:
    \begin{itemize}
    \item[]Postulate~1: The emitted laser light is purely outgoing.
    \item[]Postulate~2: The interaction of the wave with the slab is described in terms of a weak Kerr nonlinearity.
    \end{itemize}
This turns out to clarify the contribution of the mirrors and the role of their position and reflection properties in optimizing the performance of the laser.

\section{Helmholtz Equation and Its Scattering Solutions}
\label{Sec2}

Consider a homogeneous slab of thickness $L$ placed between two mirrors as shown in Fig.~\ref{fig1}. We use a coordinate system where the slab's faces are parallel to the $x$-$y$ plane and given by $z=0$ and $z=L$.

Let $\mu_1$ and $\mu_2$ label the mirrors located to left and right of the slab, respectively. For definiteness we identify these with the intervals on the $z$-axis that give the  position of the mirrors;
    \begin{align*}
    &\mu_1:=\{z|-a-t_1\leq z\leq -a\},
    &&\mu_2:=\{z|~b\leq z\leq b+t_2\},
    \end{align*}
where $a$ and $b$ are positive real parameters and $t_j$ is the thickness of $\mu_j$. We characterize the properties of the mirrors $\mu_j$ in terms of their transfer matrix:
     \begin{align}
      &\textbf{M}_{j}:=\frac{1}{T_{j}}
      \left[\begin{array}{cc}
      T_{j}^{2}- R_{j}^{r}R_{j}^{l} & R_{j}^{r} \\
      -R_{j}^{l} & 1 \\\end{array}
      \right], \notag
      \end{align}
where $R_j^{l/r}$ and $T_j$ are the complex left/right reflection and transmission amplitudes of $\mu_j$, \cite{prl-2009}. We assume that the mirrors are not active components of our system, so that their reflection and transmission coefficients, $|R^{l/r}_j|^2$ and $|T_j|^2$, add up to unity;
    \be
    |R^{l/r}_j|^2+|T_j|^2=1.
    \label{unitary}
    \ee

Suppose that the slab is made out of nonmagnetic weakly nonlinear active material, and that the system interacts with time-harmonic electromagnetic waves whose electric and magnetic fields have the form $e^{-ickt}{\vec E}({\vec r})$ and $e^{-ick t}{\vec H}({\vec r})$, respectively. Here $c$ is speed of light in vacuum and $k$ is the wavenumber. We describe the electromagnetic properties of the slab using its permittivity:
   \be
    \epsilon({\vec r})=\epsilon_0\left[\fn^2+\sigma|{\vec E}({\vec r})|^2\right],
    \label{kerr}
    \ee
where $\epsilon_0$ is the permittivity of the vacuum, $\fn$ is the complex refractive index of the slab in the absence of the nonlinearity, and $\sigma$ is the Kerr coefficient.

Throughout this article we confine our attention to the study of transverse electric (TE) waves. The electric field ${\vec E}({\vec r})$ and the wave vector $\vec k$ for a TE wave have the form
    \begin{align}
    &{\vec E}({\vec r})=e^{ik_{x}x} \sE(z){\vec e}_y, &&\vec k=k_x{\vec e}_x+k_z{\vec e}_z,
    \label{TE-defn}
    \end{align}
where ${\vec e}_x$, ${\vec e}_y$, and ${\vec e}_z$ are respectively the unit vectors along the $x$-, $y$-, and $z$-axes,
    \begin{align}
    &k_x:=k\sin\theta, &&k_z:=k\cos\theta,
    \label{kx-kz=}
    \end{align}
$\theta$ is the incidence angle depicted in Fig.~\ref{fig1}, and $\sE(z)$ is a function whose form is determined by Maxwell's equations \cite{jackson}. In view of (\ref{kerr}) -- (\ref{kx-kz=}), these reduce to
    \begin{align}
    &{\vec H}({{\vec r}}) = i (k Z_{0})^{-1}e^{ik_x x}  \left[\sE'(z){\vec e}_x-ik_x\sE(z){\vec e}_z\right],
    \label{TE-H=}\\[3pt]
    &\sE''(z) +k^2[\hat\epsilon(z)-\sin^2\theta]\sE(z) = 0,
    \label{TE-E=}
    \end{align}
where $Z_{0}$ is the vacuum impedance, $z\notin\mu_j$, and
    \be
    \hat\epsilon(z)=\left\{\begin{array}{cc}
    \fn^2+\sigma|\sE(z)|^2 & {\rm for}~ 0\leq z\leq L,\\[3pt]
    1 & {\rm otherwise} ,
    \end{array}\right.
    \label{e1}
    \ee
is the relative permittivity of our system outside the mirrors.

Next, we introduce
    \be
    \begin{aligned}
    &\mathbf{z}:=\frac{z}{L}, &&\fK:=Lk_z=kL\cos\theta,\\
    &\mathbf{a}:=\frac{a}{L}, &&\mathbf{b}:=\frac{b}{L},\quad\quad \bt_j:=\frac{t_j}{L},
    \\
    &\gamma:=-\sigma k^2L^2,
    &&\tilde\fn:=\sec\theta\sqrt{\fn^2-\sin^2\theta}.
    \end{aligned}
    \label{scaled}
    \ee
Then according to (\ref{TE-E=}) and (\ref{e1}),
    \be
    \sE(L\bz)=\left\{\begin{array}{ccc}
    A_{1}e^{i\fK\,  {\bz}}+B_{1}e^{-i\fK\, {\bz}}&{\rm for}& {\bz}< -\mathbf{a}-\bt_1,\\[3pt]
    A_{2}e^{i\fK\,  {\bz}}+B_{2}e^{-i\fK\, {\bz}}&{\rm for}& {\bz}\in [-\mathbf{a},0),\\[3pt]
    \zeta(\bz)&{\rm for}& {\bz}\in [0,1],\\[3pt]
    A_{4}e^{i\fK\,  {\bz}}+B_{4}e^{-i\fK\, {\bz}}&{\rm for}& {\bz}\in (1,\mathbf{b}],\\[3pt]
    A_{5}e^{i\fK\, {\bz}}+B_{5}e^{-i\fK\, {\bz}}&{\rm for}& {\bz}> \mathbf{b}+\bt_2,
    \end{array}\right.
    \label{e02}
    \ee
where $A_i$ and $B_i$ are complex coefficients, and  $\zeta:[0,1]\to\C$ solves the nonlinear Schr\"odinger equation:
    \be
    -\zeta''(\mathbf{z})+\fK^{2}(1-\tilde\fn^2)\zeta(\mathbf{z})+\gamma \vert\zeta(\mathbf{z})\vert^2\zeta(\mathbf{z})=\fK^{2}\zeta(\mathbf{z}),
    \label{zeta-eqn}
    \ee
for $\mathbf{z}\in(0,1)$.

The coefficients $A_i$ and $B_i$ are restricted by the standard electromagnetic interface conditions which amount to the continuity of the solution (\ref{e02})  and its derivative at $\bz=0$ and $\bz=1$ as well as the matching conditions given by the transfer matrix of the mirrors. We can write these conditions in the form:
    \begin{align}
    &\begin{aligned}
    &A_{2} = -\frac{iG_{+} (0)}{2\fK},  &&~~~~B_{2} = \frac{iG_{-} (0)}{2\fK},\\
    &A_{4} = -\frac{i\,e^{-i\fK}G_{+} (1)}{2\fK}, &&~~~~ B_{4} =
    \frac{i\,e^{i\fK}G_{-} (1)}{2\fK},
    \end{aligned}\label{e2}\\
    &\begin{aligned}
    &\left[\begin{array}{c}
        A_{2} \\
        B_{2} \end{array}\right] = \textbf{M}_{1}
        \left[\begin{array}{c}
        A_{1} \\
        B_{1} \\
      \end{array}\right],
      && \left[\begin{array}{c}
        A_{5} \\
        B_{5}\end{array}\right] = \textbf{M}_{2}
        \left[\begin{array}{c}
        A_{4} \\
        B_{4}\end{array}\right],
       \end{aligned}
        \label{mirrormatching}
    \end{align}
where
     \be
     G_{\pm} (\mathbf{z}) :=  \zeta'(\bz) \pm i\fK\,\zeta(\bz)
     \label{e3}.
     \ee

For a left-incident wave, $B_{5} = 0$, and we can use (\ref{e2}) and (\ref{mirrormatching}) to compute
    \begin{align}
    &R^{l}:=\frac{B_{1}}{A_{1}}
    =\frac{R_{1}^{l}G_{+}(0)-\left(T_{1}^{2}-R_{1}^{l} R_{1}^{r}\right)
    G_{-}(0)}{ G_{+}(0)+R_{1}^{r}G_{-}(0)},
    \label{RL=}\\
    &T^{l}:= \frac{A_{5}}{A_{1}} = \frac{T_{1}T_{2}e^{-i\fK}G_{+}(1)}{G_{+}(0) + R_{1}^{r}G_{-}(0) }.
    \label{TL=}
    \end{align}
Similarly, for a right-incident wave $A_1=0$, and (\ref{e2}) and (\ref{mirrormatching}) imply
    \begin{align}
    &R^{r}:= \frac{R^r_2e^{2i\fK}G_-(1)-\left(T_2^2-R^l_2R^r_2\right)G_+(1)}{e^{2i\fK}G_-(1)+R^l_2G_+(1)},
    \label{RR=}\\
    &T^{r}:= -\frac{T_2 e^{i\fK}}{T_1}\left[\frac{R_1^lG_+(0)-\left(T_1^2-R_1^lR_1^r\right)G_+(0)}{
    e^{2i\fK}G_-(1)+R_2^lG_+(1)}\right].
    \label{TR=}
    \end{align}

\section{Determination of Spectral Singularities}

Spectral singularities associated with the left-incident waves correspond to the poles of $R^{l}$ and $T^{l}$, \cite{prl-2009,prl-2013}. According to (\ref{RL=}) and (\ref{TL=}), they are characterized by the condition:
       \be
       G_{+}(0)+G_{-}(0)R_{1}^{r}=0.
       \label{ss}
       \ee
In view of (\ref{e3}), this is equivalent to
    \be
       \zeta'(0) + i\fK_{R_1^r}\,\zeta (0) = 0,
       \label{ss2}
       \ee
where $\fK_{R_1^r} := \fK \left(\frac{1-R_{1}^{r}}{1+R_{1}^{r}}\right)$. If we remove the first mirror,  $R_{1}^{r}=0$ and $\fK_{R_1^r} = \fK$.

Equation~(\ref{ss2}) reduces the problem of finding spectral singularities to solving the nonlinear Schr\"odinger equation (\ref{zeta-eqn}).  Assuming that $\gamma\ll 1$, we can use first-order perturbation theory for this purpose. This means that we seek for a solution of the form
      \be
      \zeta(\bz) \simeq \zeta_{0}(\bz) + \gamma \zeta_{1}(\bz),
      \label{pertsoln}
      \ee
where $\zeta_{0}(\bz)$ solves the linear Helmholtz equation,
       \be
       \zeta''_{0}(\bz) + \fK^{2}\,\tilde\fn^2\, \zeta_{0}(\bz) = 0,
       \label{linearwaveeqn}
       \ee
and fulfills the same interface conditions at $\bz=0$ and $\bz=1$,
        \be
        \zeta_{1}(\bz) :=\int_{\bz_0}^{\bz} \mathcal{G} (\bz -\bz') \left|\zeta_{0}(\bz')\right|^2\, \zeta_{0}(\bz')\,d\bz' ,
        \label{zeta1}
        \ee
$\bz_0\in[0,1]$, and $\mathcal{G} (\bz-\bz') = \sin[\fK\tilde\fn(\bz-\bz')]/\fK \tilde\fn$ is the Green's function for (\ref{linearwaveeqn}).

Next, we impose the condition (\ref{ss2}) for the existence of a spectral singularity. This leads to an equation that we expand in powers of $\gamma$ and ignore the quadratic and higher order terms. We then demand that the zero- and first-order terms vanish separately. The vanishing of the zero-order term determines the linear spectral singularities. We use subscript zero to identify the parameters associated with the emergence of the latter, e.g. $\fn_0$, $\tilde\fn_0$, $k_0$, and $\fK_0$ are the values of $\fn$, $\tilde\fn$, $k$, and $\fK$ at which a linear spectral singularity exists. The values of these quantities for which the nonlinear spectral singularities arise have the form
    \begin{align}
    &\fn=\fn_0+\gamma\,\fn_1,
    &&\tilde\fn=\tilde\fn_0+\gamma\,\tilde\fn_1,
    &&k=k_0+\gamma k_1,
    &&\fK=\fK_0+\gamma\,\fK_1,
    \label{expand}
    \end{align}
where $\fn_1,\tilde\fn_1,k_1$ and $\fK_1$ are the first-order nonlinear corrections.

\subsection{Linear spectral singularities and threshold gain}

To determine linear spectral singularities we set $\fn=\fn_0$, $\tilde\fn=\tilde\fn_0$, and $\fK=\fK_0$, and solve (\ref{linearwaveeqn}). We can easily find the general solution of this equation and impose the interface and matching conditions, (\ref{e2}) and (\ref{mirrormatching}), together with $B_5=0$ to obtain
    \begin{align}
    \zeta_{0}(\bz) = \frac{A_{5}}{2\tilde\fn_0 T_{2}} \left[U_{+}\,e^{i\fK_0\tilde\fn_0(\bz-1)} + U_{-}\,e^{-i\fK_0\tilde\fn_0(\bz-1)}\right],
    \label{linearsoln2}
    \end{align}
where
    \be
    U_{\pm}:= (\tilde\fn_0 \pm 1) e^{i \fK_0} + (\tilde\fn_0 \mp 1)R_{2}^{l} e^{-i \fK_0}.
    \label{upm}
    \ee
By definition, (\ref{linearsoln2}) satisfies  the spectral singularity condition (\ref{ss2}) with
$\fn=\fn_0$ and $\fK=\fK_0$. It is not difficult to express this condition in the form
    \be
    e^{2i \fK_0\tilde\fn_0} = \frac{V_{+}}{V_{-}}\frac{U_{+}}{U_{-}},
    \label{lss}
    \ee
where
    \be
    V_{\pm}:= (\tilde\fn_0 \pm 1) + (\tilde\fn_0 \mp 1)R_{1}^{r}.
    \label{vpm}
    \ee

Next, we write Eq.~(\ref{lss}) in terms of the quantities of direct physical significance. To this end, we first denote the real and imaginary parts of $\fn$ (respectively $\fn_0$) by $\eta$ and $\kappa$ (respectively $\eta_0$ and $\kappa_0$), so that
    \begin{align}
    &\fn=\eta+i\kappa,
    &&\fn_0=\eta_0+i\kappa_0,
    \label{eta-kappa-zero}
    \end{align}
and recall that the gain coefficient of the slab and its threshold value are respectively given by \cite{Silfvast}:
    \begin{align}
    &g=-2k\kappa, && g_0=-2k_0\kappa_0.
    \label{g-zero}
    \end{align}
We can determine $g_0$ and $k_0$ by equating the absolute-value and the phase of the left- and right-hand sides of (\ref{lss}), respectively. This gives
    \bea
    g_0&=&g_0^{(s)}+g_0^{(1)}+g_0^{(2)},
    \label{th-g=1}\\
    k_0&=&\frac{\pi m-\varphi_0}{L\cos\theta\RE(\tilde\fn_0)},
    \label{th-lambda=1}
    \eea
where
    \bea
    g_0^{(s)}&:=&\frac{2\fa_0}{L}\ln\left|\frac{\tilde\fn_0+1}{\tilde\fn_0-1}\right|,
    \label{g0-s=}\\
    g_0^{(1)}&:=&\frac{\fa_0}{L}
    \ln\left|\frac{1 + \frac{\tilde\fn_0-1}{\tilde\fn_0+1}R_1^{r}}{1 + \frac{\tilde\fn_0+1}{\tilde\fn_0-1}R_1^{r}}\right|,
    \label{g0-1=}\\
    g_0^{(2)}&:=&\frac{\fa_0}{L}
    \ln\left|\frac{1 + \frac{\tilde\fn_0-1}{\tilde\fn_0+1}R_2^{l}
    e^{-2i\fK_0}}{1 + \frac{\tilde\fn_0+1}{\tilde\fn_0-1} R_2^{l}e^{-2i\fK_0}}\right|,
    \label{g0-2=}\\
    \fa_0&:=&\frac{\IM(\fn_{0})}{\cos\theta\,\IM(\tilde\fn_{0})},
    \label{fa0=}
    \eea
$m=1,2,3,\ldots$ is a mode number, and $\varphi_0$ is the phase angle of $V_{+} U_{+}/V_{-} U_{-}$. $g_0^{(s)}$ and $g_0^{(j)}$ respectively give the contribution of the slab and the mirror $\mu_j$ to the threshold gain. In the absence of the mirrors, $R_1^r=R_2^l=0$, $g_0^{(1)}=g_0^{(2)}=0$, and we find $g_0=g_0^{(s)}$. This coincides with the formula obtained in \cite{pra-2015a} for the threshold gain coefficient pertaining the TE modes of a mirrorless slab.

In Appendix~A, we give the explicit form of $g_0^{(s)}$, $g_0^{(1)}$, and $g_0^{(2)}$ for a slab made of a typical high-gain material with a thickness much larger than the wavelength of the emitted wave, where $|\kappa_0|\ll \eta_0-1\ll k_0L$.

As expected the presence of mirrors can have drastic effects on the threshold gain. In order to arrive at a quantitative description of these effects, first we express the reflection and transmission amplitudes, $R_j^{l/r}$ and $T_j$, of the mirrors in terms of the reflection and transmission amplitudes, $\sR_j^{l/r}$ and $\sT_j$, of identical mirrors that are adjacent to the slab boundaries. Because under a translation $z\to z-d$, $R_j^{l/r}$ and $T_j$ transform according to: $R_j^{r}\to e^{-2idk_z}R_j^{r}$, $R_j^{l}\to e^{2idk_z}R_j^{l}$, and $T_j\to T_j$, \cite{pra-2014b}, for $k=k_0$ we have
    \begin{align}
    & R_1^{r} = e^{2i\bba\, \fK_0} \sR_1^{r},
    && R_2^{l} = e^{2i\bb'\fK_0} \sR_2^{l}, && T_j = \sT_j,
    \label{reflections-transmissions}
    \end{align}
where $\bba := a/L$ and $\bb' := \bb - 1 = b/L -1$. As noted in Ref.~\cite{Haus}, we can take $\sR^{r}_1$ and $\sR^{l}_2$ real and $\sT_j$ purely imaginary, so that
    \begin{align}
    &T_1=i\sqrt{1-{\sR^{r}_1}^2},
    && T_2=i\sqrt{1-{\sR^{l}_2}^2}.
    \label{T-R=}
    \end{align}

\begin{figure}
    \begin{center}
    \includegraphics[scale=.70]{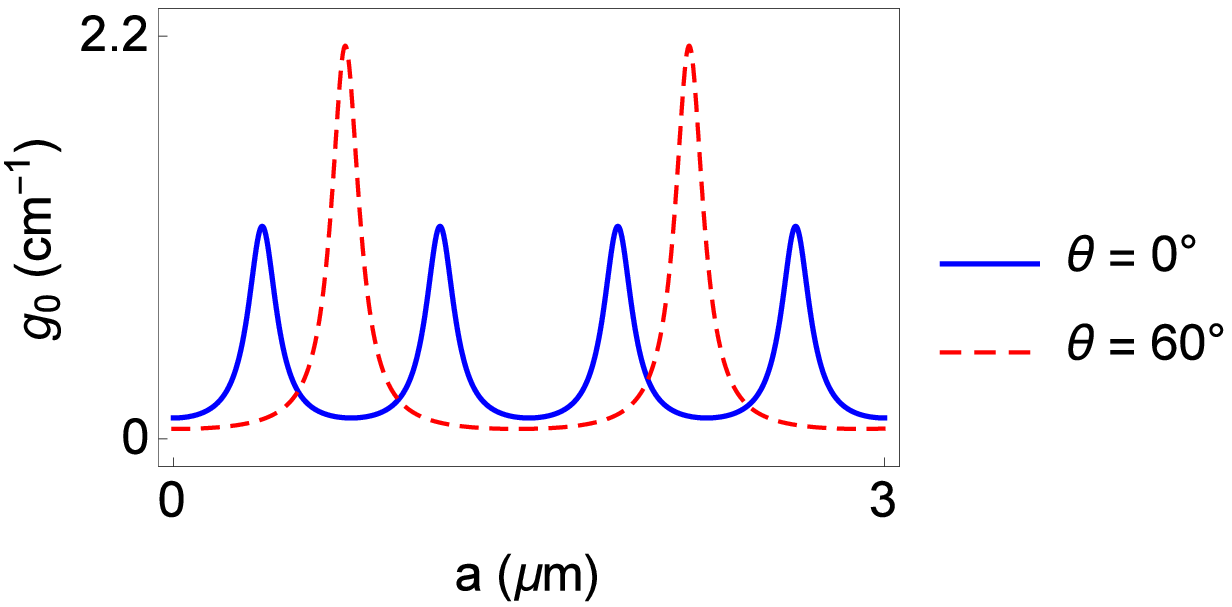}~~~~
    \includegraphics[scale=.70]{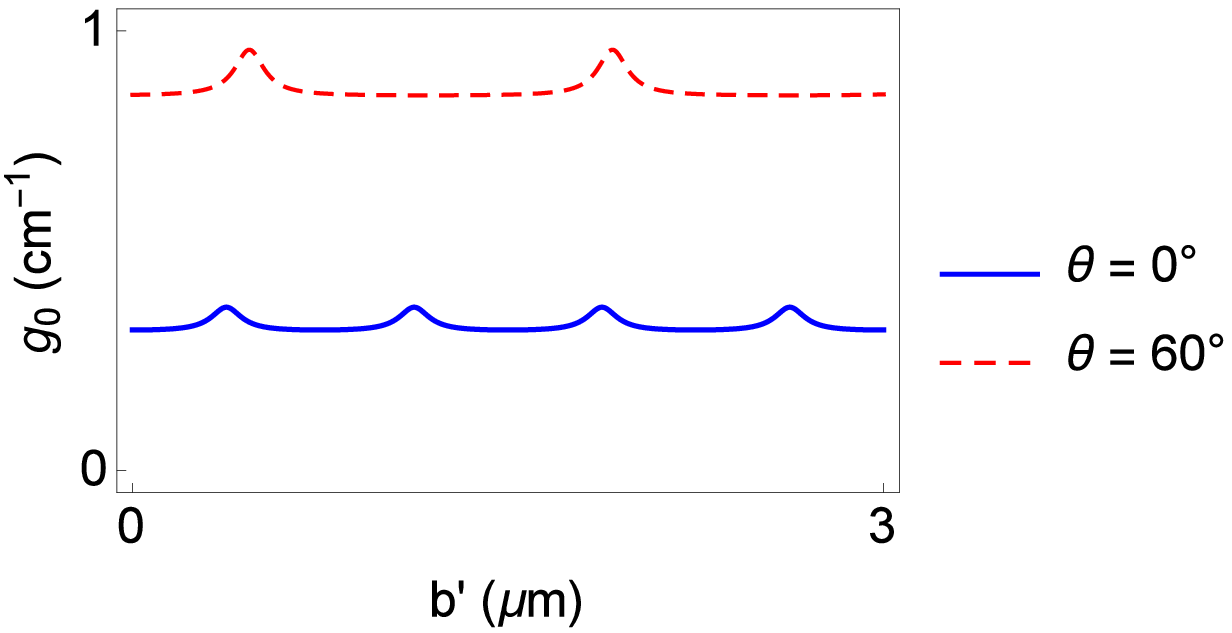}
    \caption{(Color online) Plots of the threshold gain $g_0$ for incidence angles $\theta=0^\circ$ and $60^\circ$ as a function of the distance between the slab and the mirrors, i.e., $a$ for the left mirror and $b':=b-L$ for the right mirror. The relevant physical parameters are given by (\ref{specifics}) and (\ref{specifics-mirrors}) .}
    \label{fig2}
    \end{center}
    \end{figure}

\begin{figure}
    \begin{center}
    \includegraphics[scale=.50]{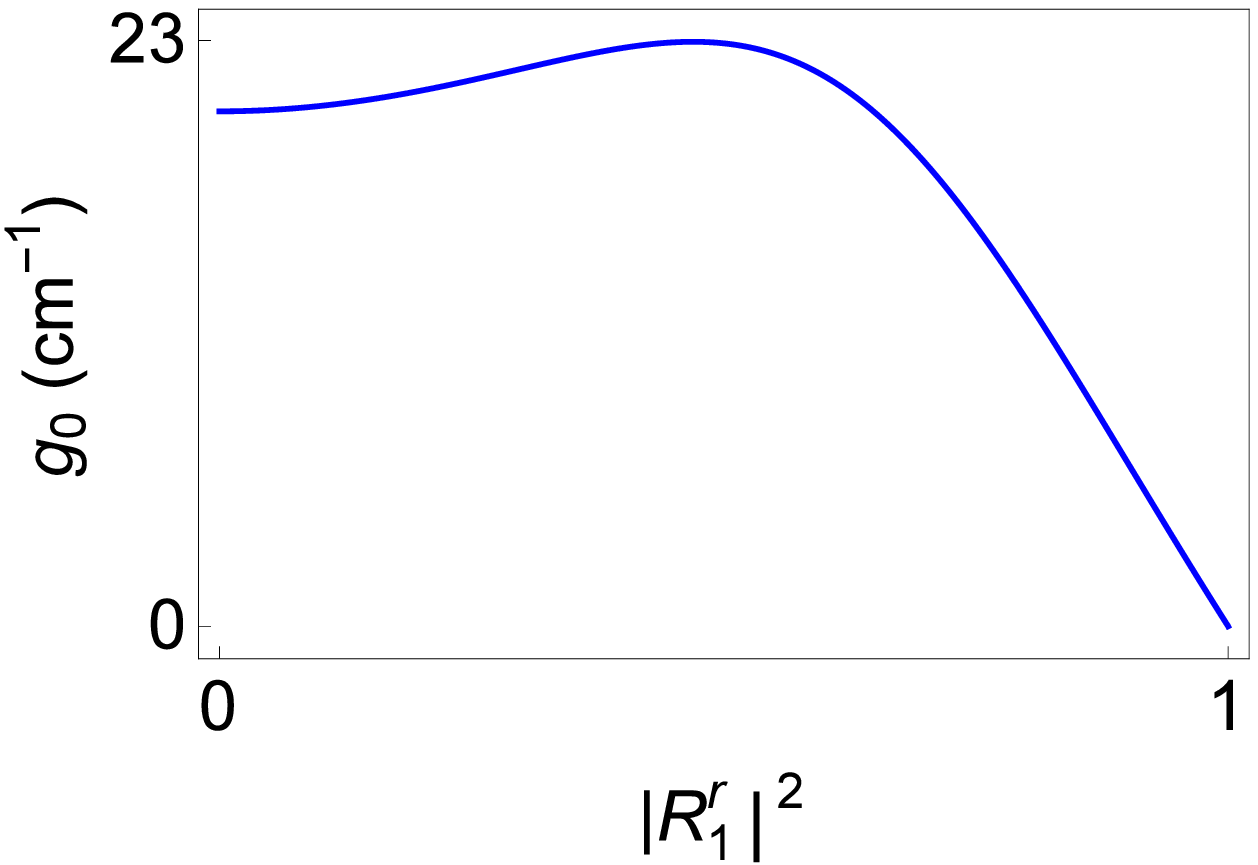}~~~~
    \includegraphics[scale=.50]{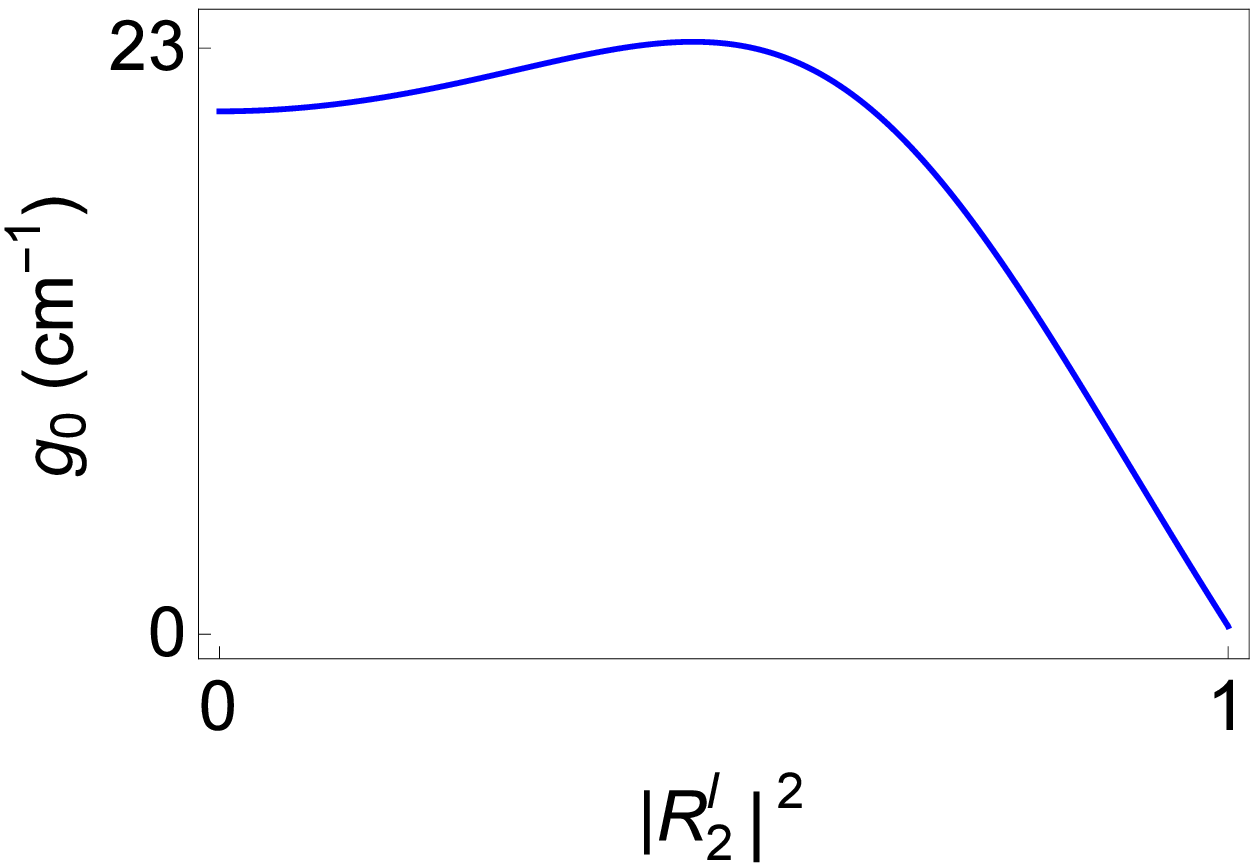}
    \caption{(Color online)  Plots of the threshold gain $g_0$ for a normally incident TE wave as a function of the reflectance of the mirrors, $|R_1^{r}|^2$ and $|R_2^{l}|^2$. To plot the left and right graphs we have respectively chosen $|R_2^{l}|^2=99.9\%$ and $|R_1^{r}|^2= 98\%$. Other relevant physical parameters are given by (\ref{specifics}).}
    \label{fig3}
    \end{center}
    \end{figure}

\begin{figure}
    \begin{center}
    \includegraphics[scale=.50]{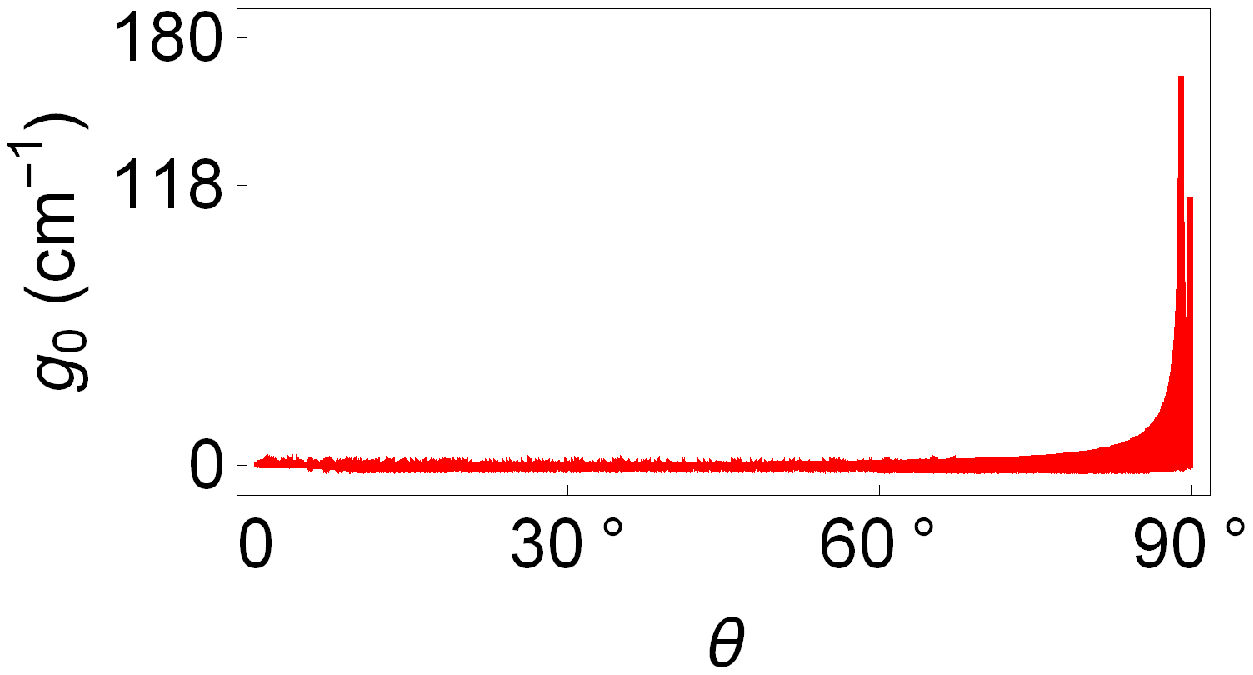}~~~
    \includegraphics[scale=.50]{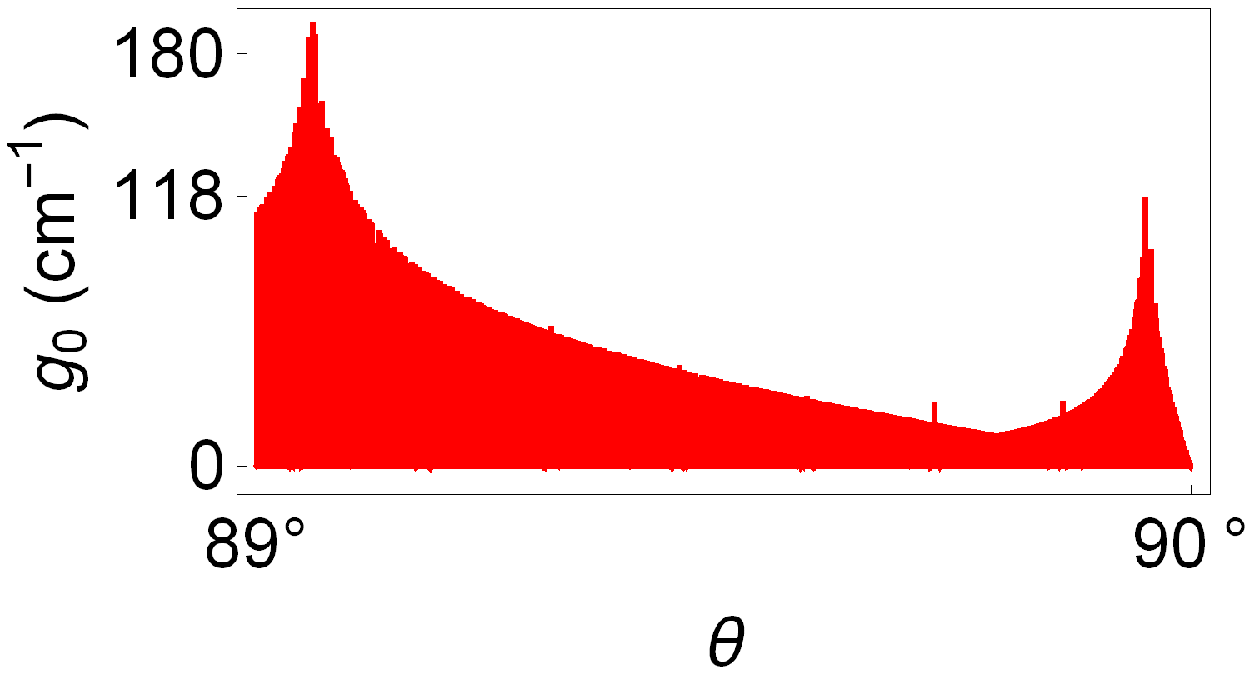}~~~\\
    \includegraphics[scale=.50]{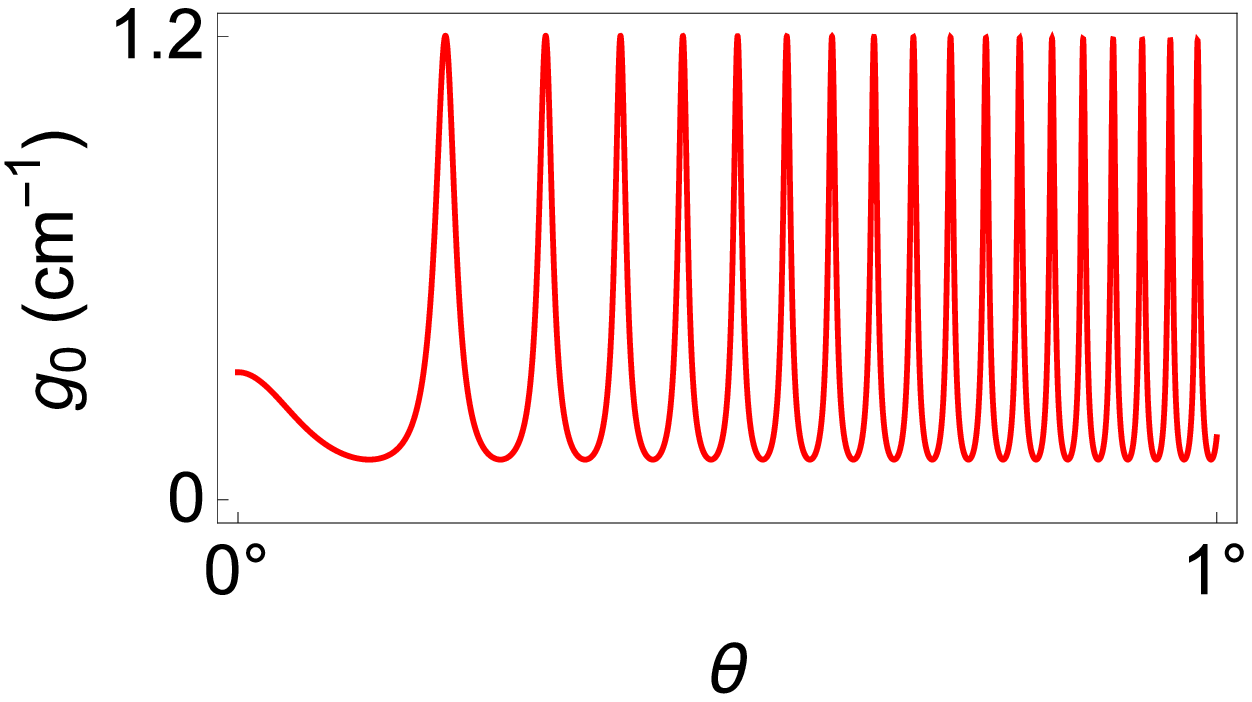}~~~~
    \includegraphics[scale=.50]{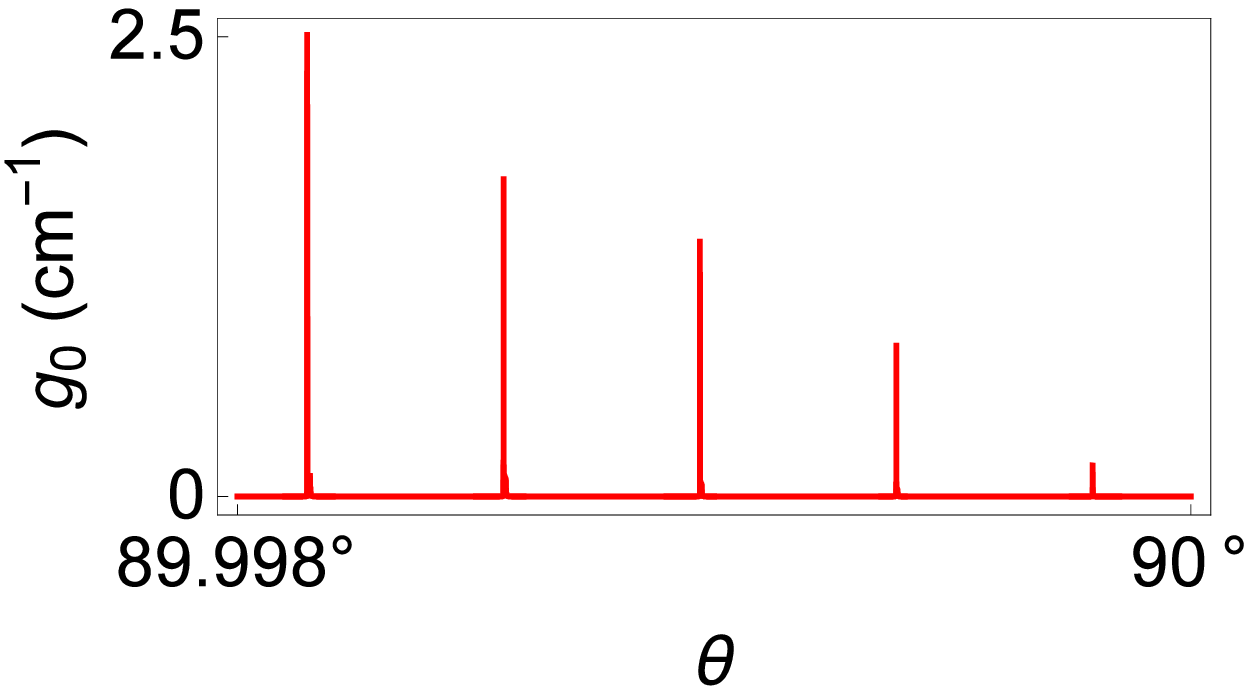}
    \caption{(Color online) Plots of the threshold gain $g_0$ as a function of $\theta_0$. The relevant physical parameters are given by (\ref{specifics}) and (\ref{specifics-mirrors}). The highest peaks are located at $\theta_0=89.083^\circ$ and $\theta_0=89.955^\circ$. They depend on the position and the reflectivity of the mirrors.}
    \label{fig4}
    \end{center}
    \end{figure}

\begin{figure}
    \begin{center}
    \includegraphics[scale=.40]{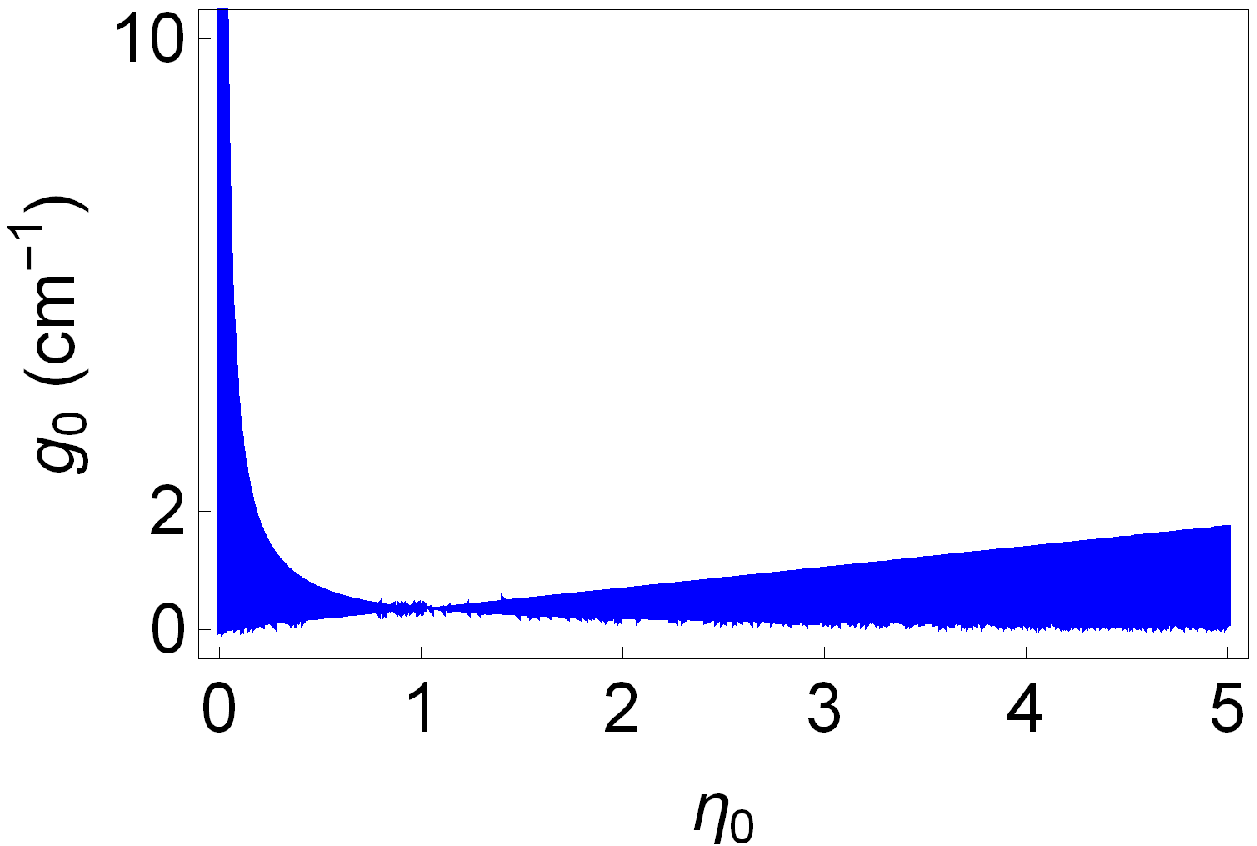}~~~
    \includegraphics[scale=.46]{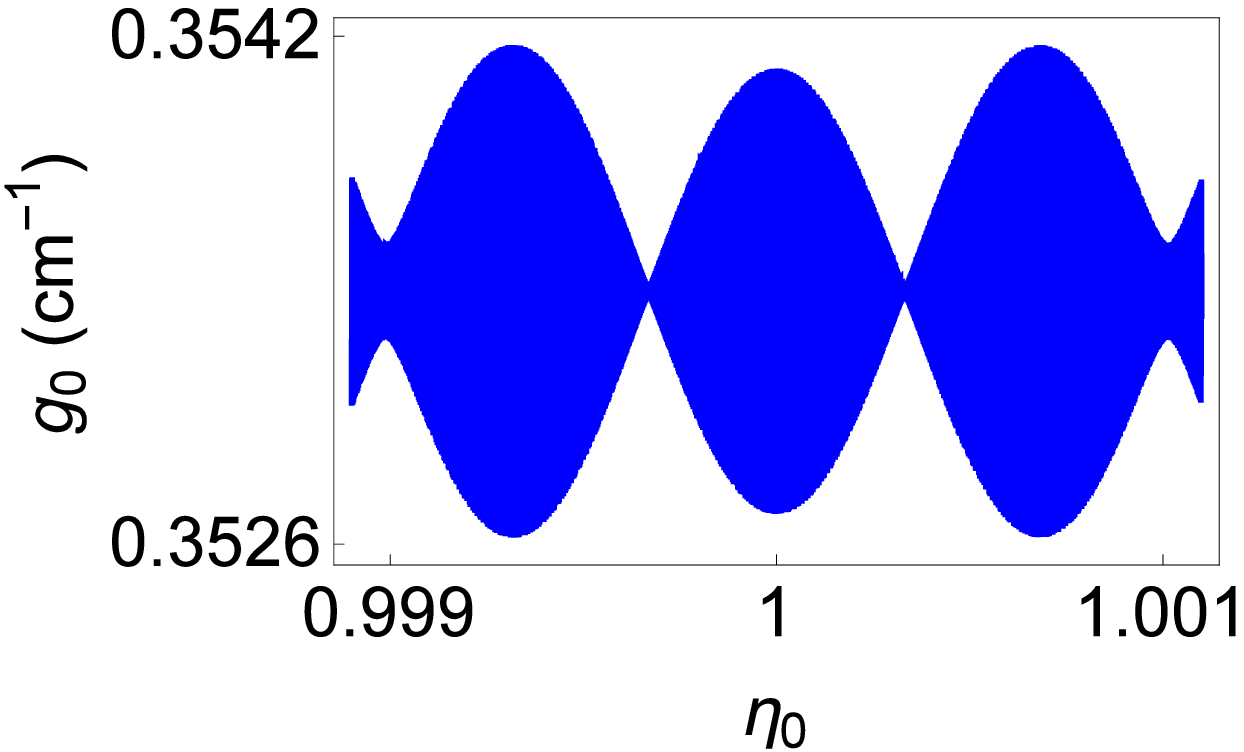}~~~
    \includegraphics[scale=.45]{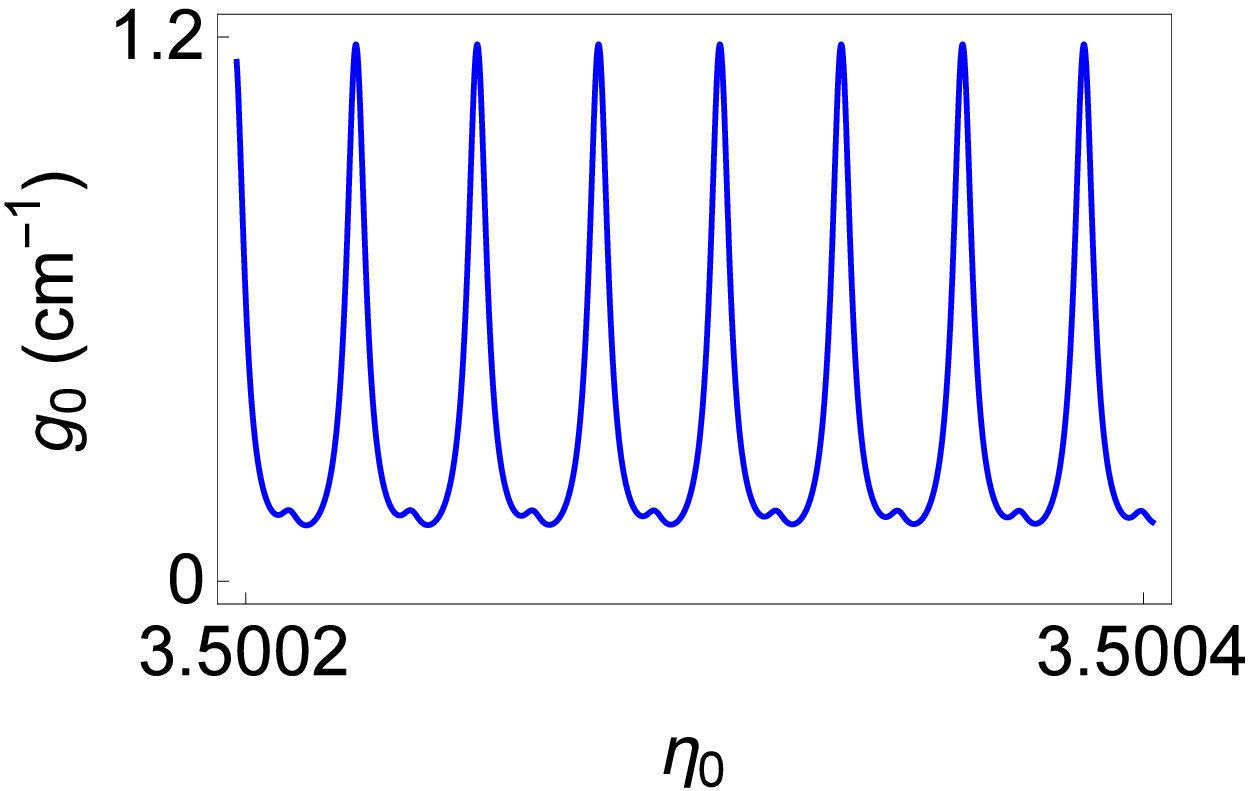}
    \caption{(Color online) Graph of the threshold gain $g_0$ as a function of the real part of the refractive index {$\eta_0$} for a normally incident TE wave. Here we have used the numerical values given by (\ref{specifics}) and (\ref{specifics-mirrors}) for the relevant physical parameters except for the value of $\eta_0$ which is the independent variable.}
    \label{fig5}
    \end{center}
    \end{figure}

Figures~\ref{fig2}-\ref{fig5} provide graphical demonstrations of the consequences of the exact expression for the threshold gain $g_0$ that is given by (\ref{th-g=1}) -- (\ref{fa0=}). To produce these graphs we have used the following values for the relevant physical parameters.
    \begin{align}
    & \eta_0=3.4,~~~~L = 300~\mu\textrm{m},~~~~m =1360,~~~~\lambda_0=1500~\textrm{nm},~~~~a = b' = 10~\textrm{cm},
    \label{specifics}\\
    & \sR_1^{r}\approx0.98995~~(|R_1^{r}|^2= 98\%),
    ~~~~~~\sR_2^{l}\approx0.99950~~(|R_2^{l}|^2= 99.9\%).
    \label{specifics-mirrors}
    \end{align}
Fig.~\ref{fig2} shows the plots of $g_0$ as a function of the distance between the slab and the mirrors. The minima of the curves correspond to the optimal positions of the mirrors.
Fig.~\ref{fig3} describes the effects of changing the reflectivity of the mirrors on $g_0$ for a normally incident TE wave ($\theta=0$). It reveals the surprising fact that, for a sufficiently poor mirror, $g_0$ is not a decreasing function of its reflectivity.
Fig.~\ref{fig4} reveals the dependence of $g_0$ on the incidence angle $\theta$. For a value of $g_0$ between its minimum and maximum values there are intervals of values of the incidence angle $\theta$ where the system can attain and exceed the threshold gain and laser oscillations begin. The two highest peaks of the $g_0$-$\theta$ graph are reminiscent of the maximum of the threshold gain curve for the TM modes of a mirrorless slab that is attained at the Brewster's angle \cite{pra-2015a}. These peaks are related to the presence of the mirrors. Their location and height depend on the reflectivity of the mirrors. Removing each of the mirrors results in the disappearance of the corresponding peak.
Fig.~\ref{fig5} demonstrates the effect of the real part of the refractive index $\eta_0$ on the threshold gain coefficient $g_0$ for a normally incident TE wave.

\subsection{Nonlinear spectral singularities and laser output intensity}

Nonlinear spectral singularities of our system are characterized by (\ref{ss}) or (\ref{ss2}). In what follows we wish to derive an explicit form of these conditions in terms of the physical parameters of the system. First, we introduce
    \be
    G(\fn,\fK) := G_{+} (0) + G_{-} (0) R_{1}^{r},
    \label{gplus}
    \ee
so that (\ref{ss}) reads $G(\fn,\fK) = 0$. We then use first-order perturbation theory in the nonlinearity parameter $\gamma$ to satisfy this relation. To do this we write
    \be
    G(\fn,\fK)= G^{(0)} (\fn, \fK) + \gamma G^{(1)} (\fn, \fK),
    \label{ss-non2}
    \ee
and assume that $G^{(\ell)} (\fn, \fK)$ are independent of $\gamma$. We then substitute $\fn=\fn_0+\gamma\fn_1$ and $\fK=\fK_0+\gamma\fK_1$ in the right-hand side of (\ref{ss-non2}), determine the terms of order $\gamma^0=1$ and $\gamma^1=\gamma$ of the resulting expression, and demand that they vanish separately. Doing this for the zero-order term yields $G^{(0)} (\fn, \fK)=0$ which by construction specifies linear spectral singularities and results in $\fn=\fn_0$ and $\fK=\fK_0$. Requiring the first-order term to vanish gives
    \be
    \fn_1\partial_{\fn_0} G^{(0)}(\fn_0,\fK_0)+\fK_1\partial_{\fK_0}G^{(0)}(\fn_0, \fK_0)
    +G^{(1)} (\fn_0, \fK_0) = 0.
    \label{nonlinearss}
    \ee
Our task is to explore the consequences of this equation.

Let $\eta_1$ and $\kappa_1$ respectively denote the real and imaginary parts of $\fn_1$, so that
    \be
    \fn_1 = \eta_1 + i \kappa_1.
    \label{fn-2=}
    \ee
Substituting this equation in (\ref{nonlinearss}) and supposing that the presence of nonlinearity does not change the value of $\eta$, i.e., setting $\eta_1=0$, we can solve (\ref{nonlinearss}) for $\kappa_1$ and $\fK_1$. This yields
    \begin{align}
    \kappa_1 &= \frac{\RE [G_{+}^{(1)}]\,\IM [\partial_{\fK_0} G_{+}^{(0)}] - \IM [G_{+}^{(1)}]\,\RE [\partial_{\fK_0} G_{+}^{(0)}]}{\IM [\partial_{\fK_0} G_{+}^{(0)}]\,\IM [\partial_{\fn_0} G_{+}^{(0)}] + \RE [\partial_{\fn_0} G_{+}^{(0)}]\,\RE [\partial_{\fK_0} G_{+}^{(0)}]},
    \label{eq1021}\\
    \fK_1 &= -\frac{\IM [G_{+}^{(1)}]\,\IM [\partial_{\fn_0} G_{+}^{(0)}] + \RE [G_{+}^{(1)}]\,\RE [\partial_{\fn_0} G_{+}^{(0)}]}{\IM [\partial_{\fK_0} G_{+}^{(0)}]\,\IM [\partial_{\fn_0} G_{+}^{(0)}] + \RE [\partial_{\fn_0} G_{+}^{(0)}]\,\RE [\partial_{\fK_0} G_{+}^{(0)}]}.
    \label{eq1022}
    \end{align}
In Appendix~B, we give explicit formulas for $\partial_{\fK_0} G_{+}^{(0)}$, $\partial_{\fn_0} G_{+}^{(0)}$, and $G_{+}^{(1)}$. Using these formulas or by direct inspection, we observe that the right-hand side of (\ref{eq1021}) and (\ref{eq1022}) are proportional to $|A_5|^2$. In other words
    \begin{align}
    &\widehat\kappa_1:=\frac{\kappa_1}{|A_5|^2},
    &&\widehat{\fK}_1:=\frac{\fK_1}{|A_5|^2},
    \label{hat-}
    \end{align}
are $A_5$-independent.

Next, we use (\ref{expand}), (\ref{g-zero}), and (\ref{fn-2=}) to show that up to the linear term in $\gamma$ the gain coefficient has the form \cite{pra-2013c}:
    \be
    g=g_0\left[1+\gamma\left(\frac{\fK_1}{\fK_0}+\frac{\kappa_1}{\kappa_0}\right)\right].
    \label{g-g0}
    \ee
Substituting (\ref{hat-}) in this relation and noting that the time-averaged intensity of the wave emitted from the right-hand face of the mirror $\mu_2$ is given by $I_r:=|A_5|^2/2$, we can show that $I_r$ satisfies (\ref{eq1}), i.e.,
    \be
    I_r=\left(\frac{g-g_0}{\sigma g_0}\right)\hat I_r,
    \label{eq1-r}
    \ee
where
    \be
    \widehat I_{r}:= -\frac{\kappa_0\cos\theta^2}{2\fK_0(\widehat\kappa_1\fK_0 + \kappa_0\widehat{\fK}_1)}.
    \label{rightintensity}
    \ee
This is the intensity slope that encodes the specific information about the output intensity of the  slab laser for waves propagating to the right.

Next, we determine the time-averaged output intensity for waves propagating to the left, i.e., $I_l:=|B_1|^2/2$. To do this we first recall that spectral singularities correspond to purely outgoing waves. In our case, this is equivalent to $A_1=B_5=0$. With the help of this relation and Eqs.~(\ref{e2}), (\ref{mirrormatching}), and (\ref{lss}), we find
      \be
      A_5=\left(\frac{T_{2}\,V_{-}\,e^{i\fK_0\tilde{\fn}_0}}{T_1\,U_{+}}\right)B_{1}.\notag
      \ee
We can use this equation together with (\ref{hat-}) to express the $\fK_1$ and $\kappa_1$ appearing on the right-hand side of (\ref{g-g0}) in terms of $B_1$. This in turn implies
     \be
     I_{l} = \frac{\left|B_1\right|^2}{2} = \left(\frac{g-g_0}{\sigma g_0}\right) \widehat I_{l},
     \label{eq1-l}
     \ee
where
    \be
    \widehat I_{l} =\left|\frac{U_{-}}{V_{+}}\right|^2
    \sqrt{\frac{1-|R_1^r|^2}{1-|R^l_2|^2}}\;e^{-2\fK_0\IM(\tilde{\fn}_0)}\;\widehat I_{r}.
    \label{leftintensity}
    \ee

We can obtain the explicit form of the intensity slopes $\widehat I_{r/l}$ by substituting the expression for $\partial_{\fK_0} G_{+}^{(0)}$, $\partial_{\fn_0} G_{+}^{(0)}$, and $G_{+}^{(1)}$, which we give in Appendix~B, in (\ref{eq1021}) and (\ref{eq1022}) and using the result in (\ref{hat-}), (\ref{rightintensity}), and (\ref{leftintensity}). This leads to extremely lengthy formulas for $\widehat I_{r/l}$ which we do not include here. We suffice to provide a graphical examination of their physical implications. Here we employ (\ref{specifics}) and (\ref{specifics-mirrors}) and set $\left|T_1\right|^2 = 2\%$ and $\left|T_2\right|^2 = 0.1\%$ which follow from (\ref{specifics-mirrors}) and (\ref{T-R=}).

\begin{figure}
    \begin{center}
    \includegraphics[scale=.60]{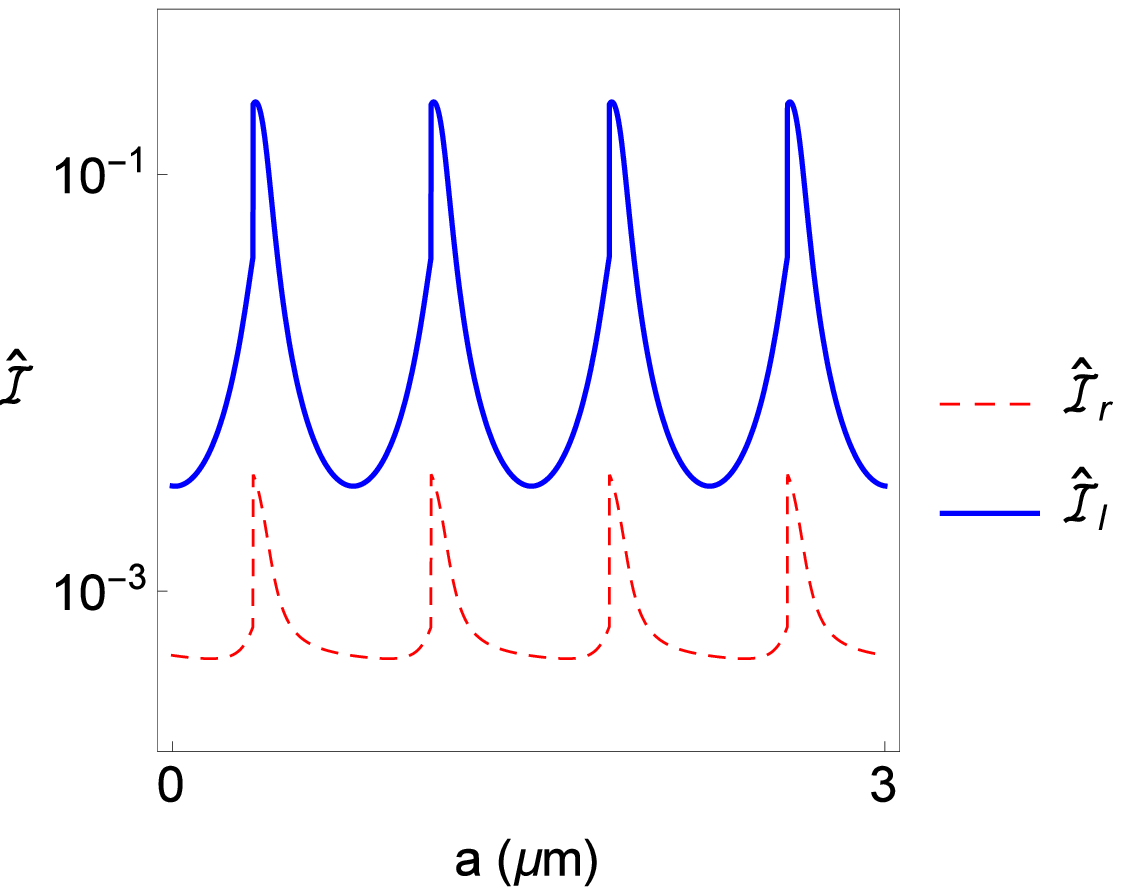}~~~
    \includegraphics[scale=.60]{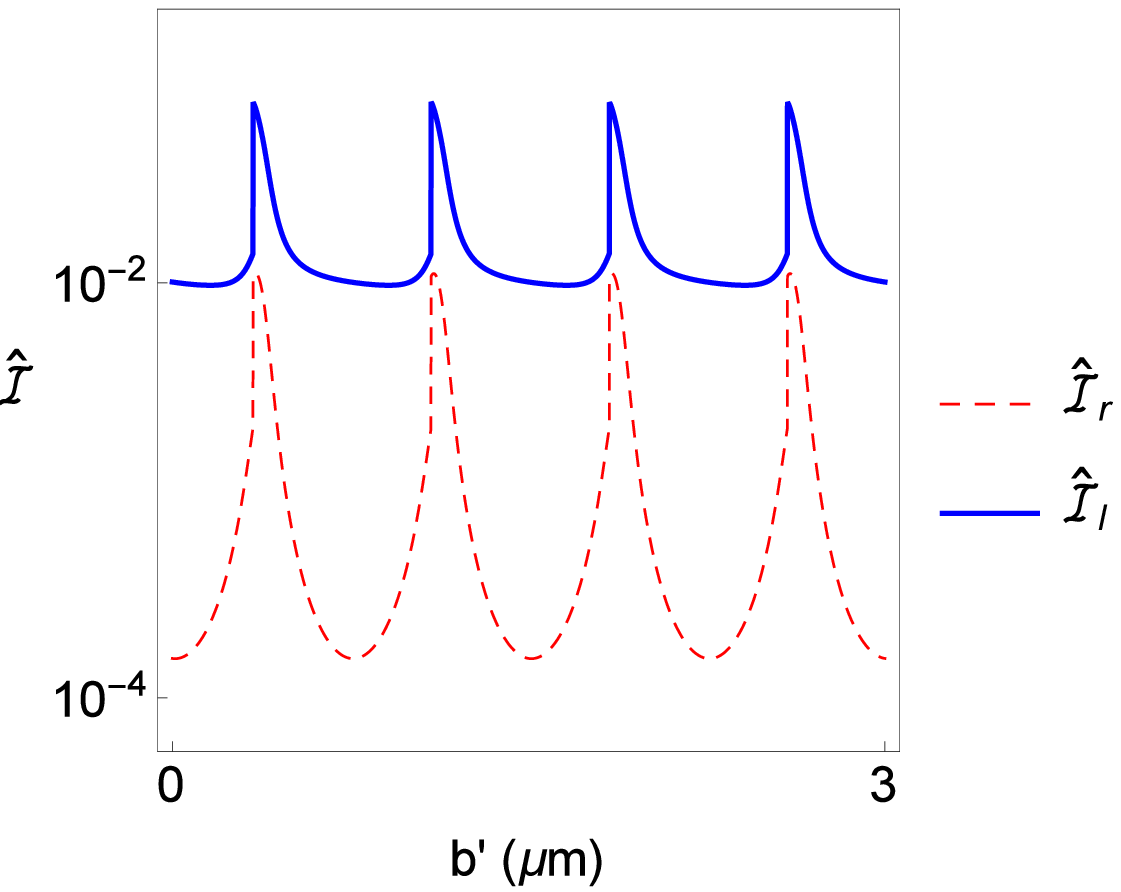}
    \caption{(Color online) Logarithmic plots of $\widehat I_{l}$ and $\widehat I_{r}$ as a function of the distance of the mirrors from the slab, i.e., $a$ with $b'=10~{\rm cm}$ (on the left) and $b'$ with $a=10~{\rm cm}$ (on the right), for $\theta = 0^{\circ}$. Values of the other relevant physical parameters are given by (\ref{specifics}) and (\ref{specifics-mirrors}).}
    \label{fignl1}
    \end{center}
    \end{figure}

\begin{figure}
    \begin{center}
    \includegraphics[scale=.60]{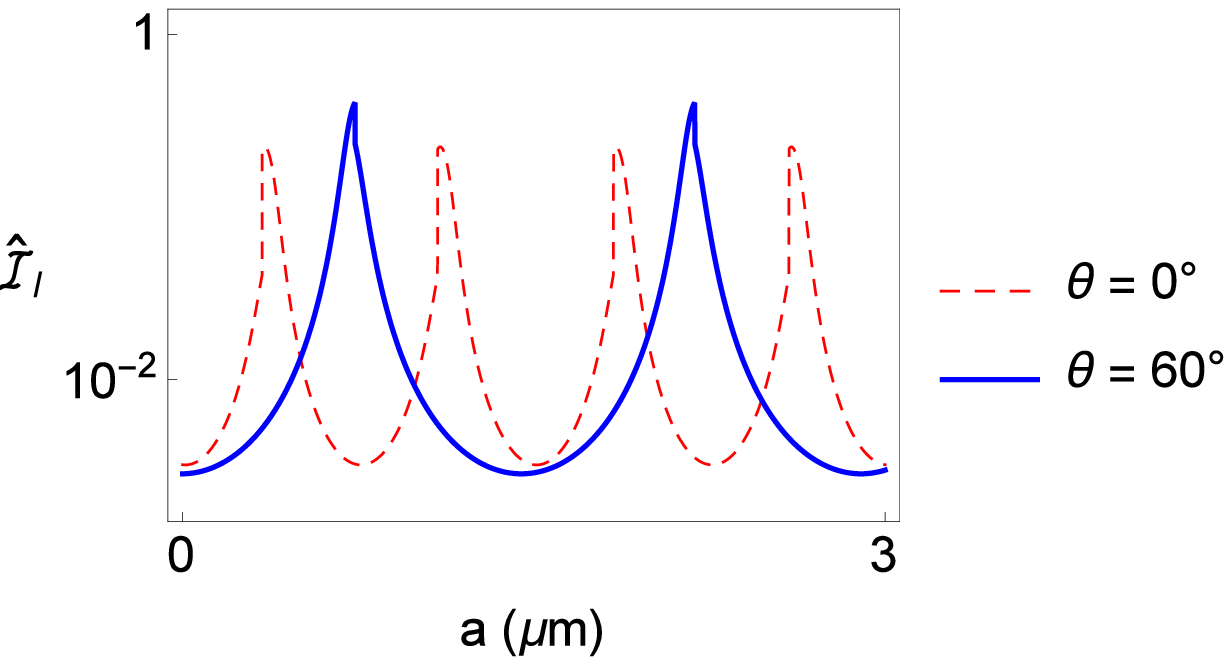}~~~
    \includegraphics[scale=.60]{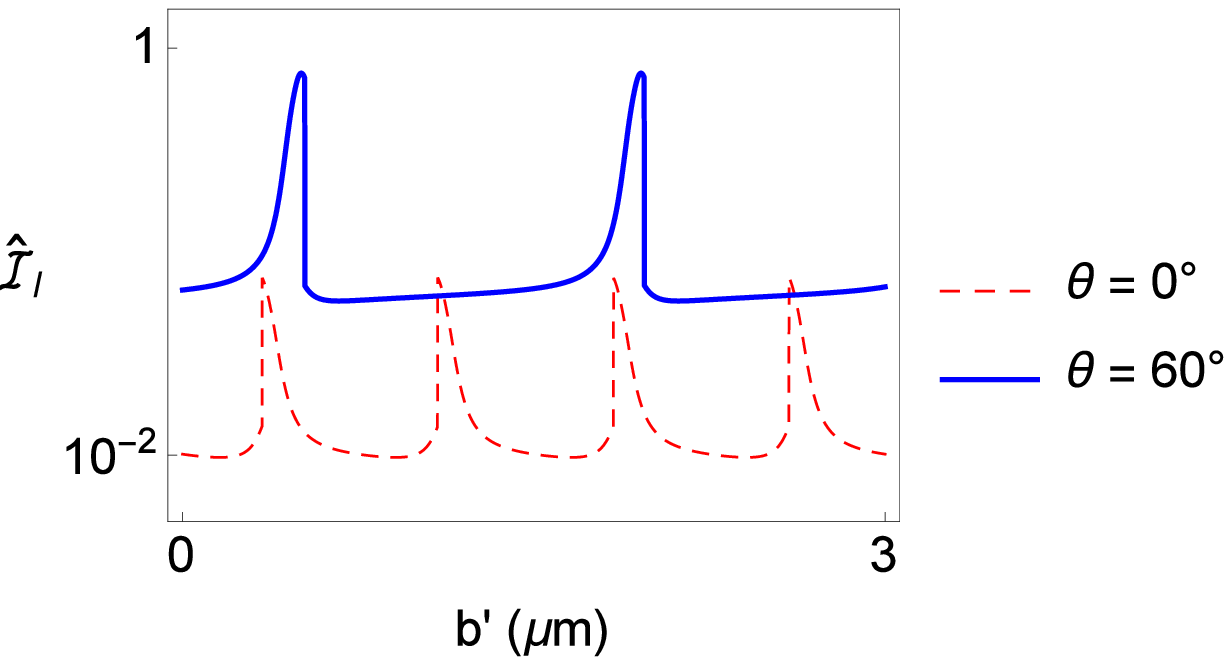}
    \caption{(Color online) Logarithmic plots of $\widehat I_{l}$ as functions of the distance of the mirrors from the slab, i.e., $a$ with $b'=10~{\rm cm}$ (on the left panel) and $b'$ with $a=10~{\rm cm}$ (on the right panel), for $\theta = 0^{\circ}$ and $60^{\circ}$. Values of the other relevant physical parameters are given by (\ref{specifics}) and (\ref{specifics-mirrors}).}
    \label{fign2}
    \end{center}
    \end{figure}

Fig.~\ref{fignl1} shows the dependence of $\widehat I_{l}$ and $\widehat I_{r}$ on the position of the mirrors for right-incident TE waves ($\theta = 0^{\circ}$). As expected, the output intensity for the waves emitted from the left mirror, i.e., $\mu_1$,  is much larger than the one from the right mirror. This is simply because $|T_1|^2=20|T_2|^2$. The peaks correspond to the optimal positions of the mirrors.
We have checked that the general behavior depicted in Fig.~\ref{fignl1} applies also for oblique TE waves. The only difference is that the distance between the adjacent peaks depends on the value of the incidence angle. This is shown in Fig.~\ref{fign2} for $\widehat I_{l}$.

\begin{figure}
    \begin{center}
    \includegraphics[scale=.60]{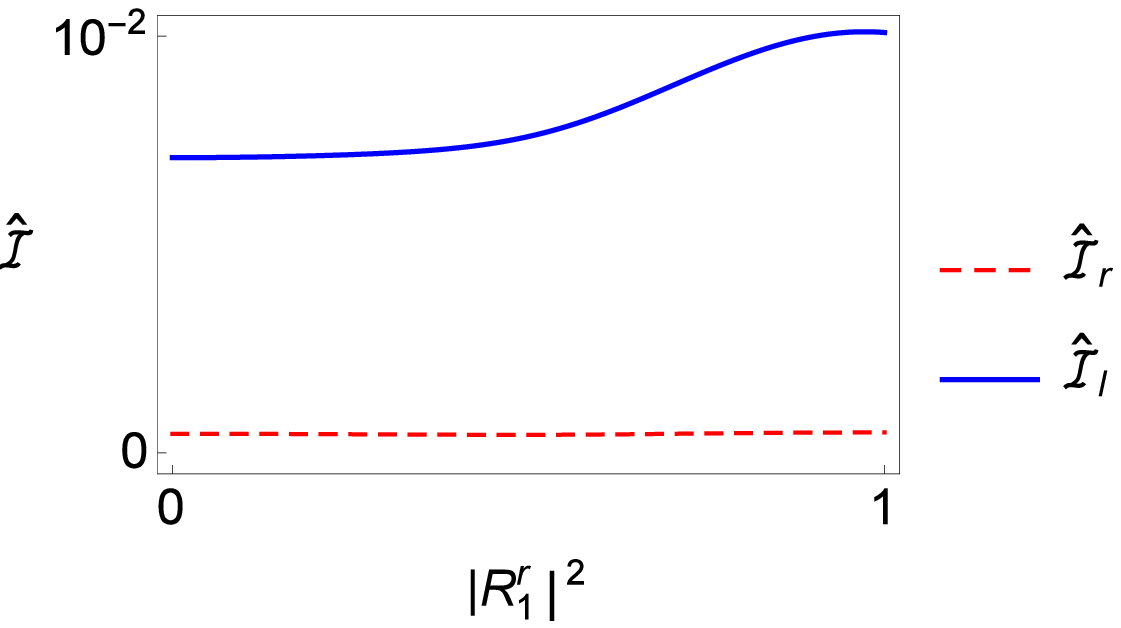}~~~
    \includegraphics[scale=.60]{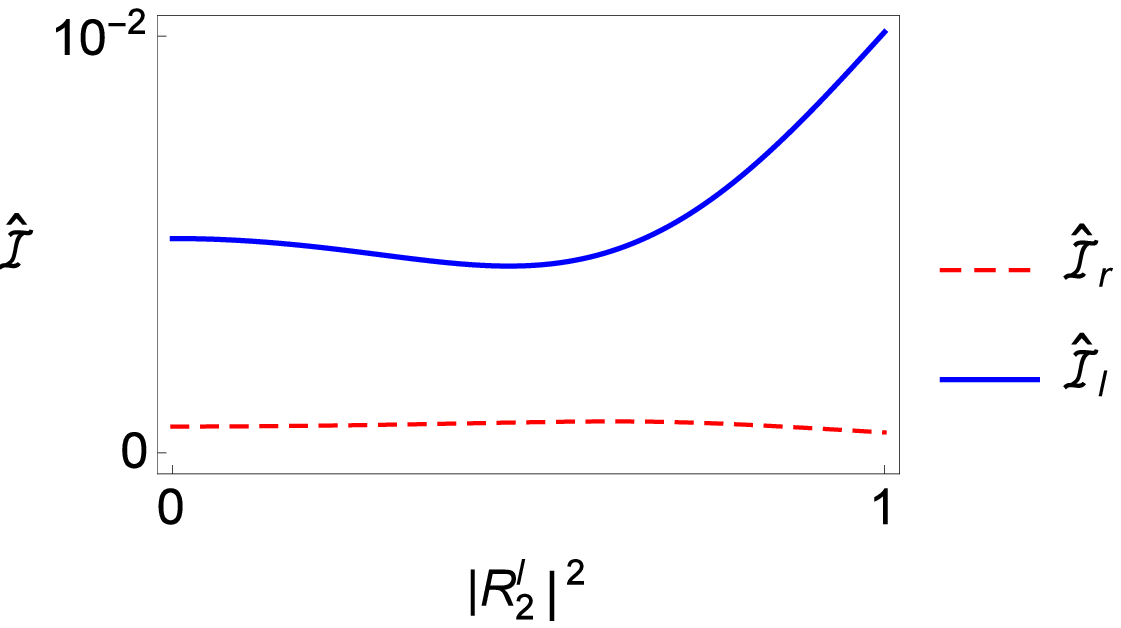}
    \caption{(Color online) Logarithmic plots of $\widehat I_{l}$ and $\widehat I_{r}$ as a function of $|R^r_1|^2$ with $|R^l_2|^2=99.9\%$ (on the left) and $|R^l_2|^2$ with $|R^r_1|^2=98\%$ (on the right) for $\theta = 0^{\circ}$. Values of the other relevant physical parameters are given by (\ref{specifics}) and (\ref{specifics-mirrors}).}
    \label{fignl2}
    \end{center}
    \end{figure}

Fig.~\ref{fignl2} reveals the effect of the change in the reflectance of the mirrors on the output intensity of emitted waves. The fact that $\widehat I_{l}>\widehat I_{r}$ is clearly seen from this figure.



\begin{figure}
    \begin{center}
    \includegraphics[scale=.45]{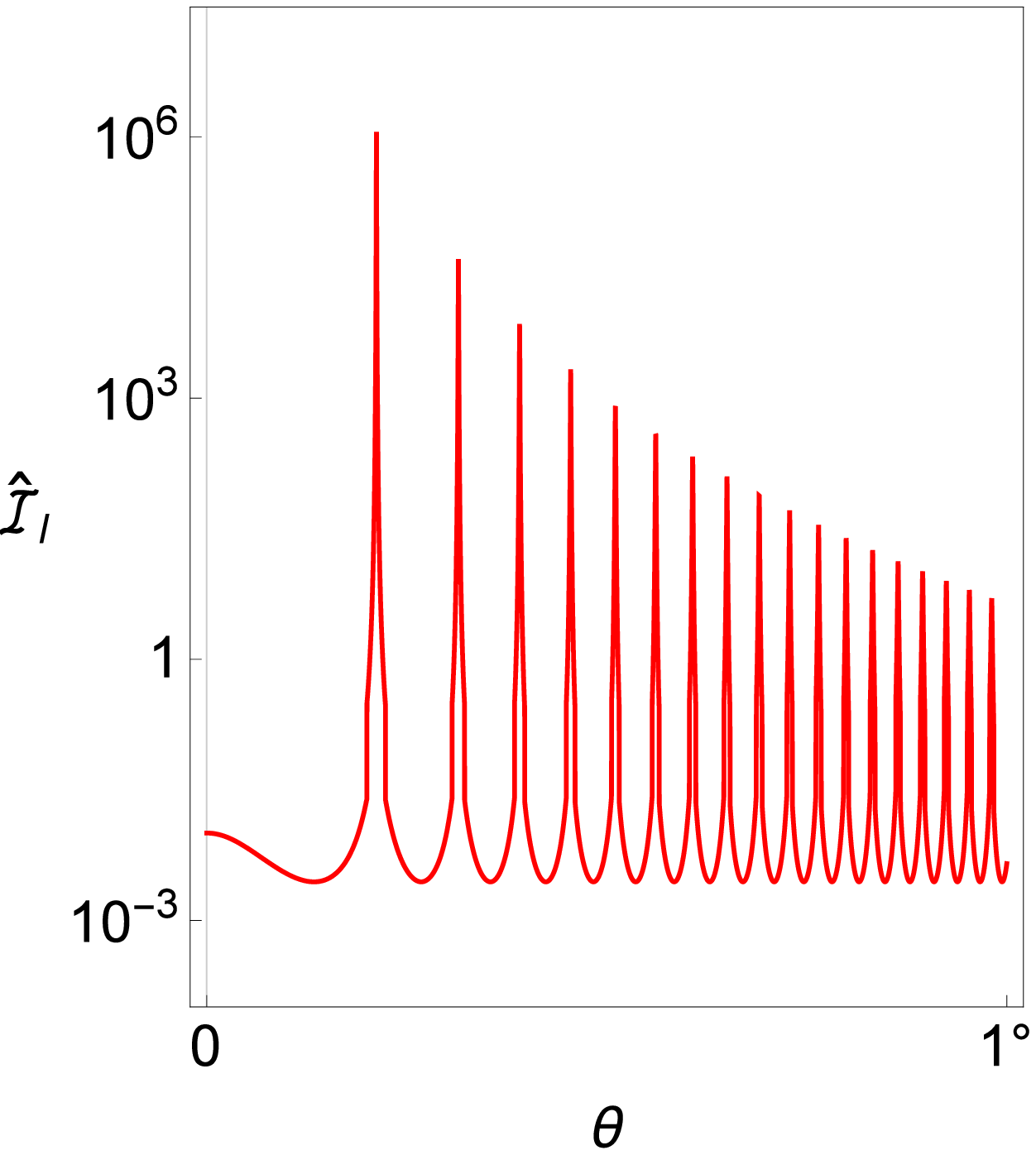}~~~
    \includegraphics[scale=.45]{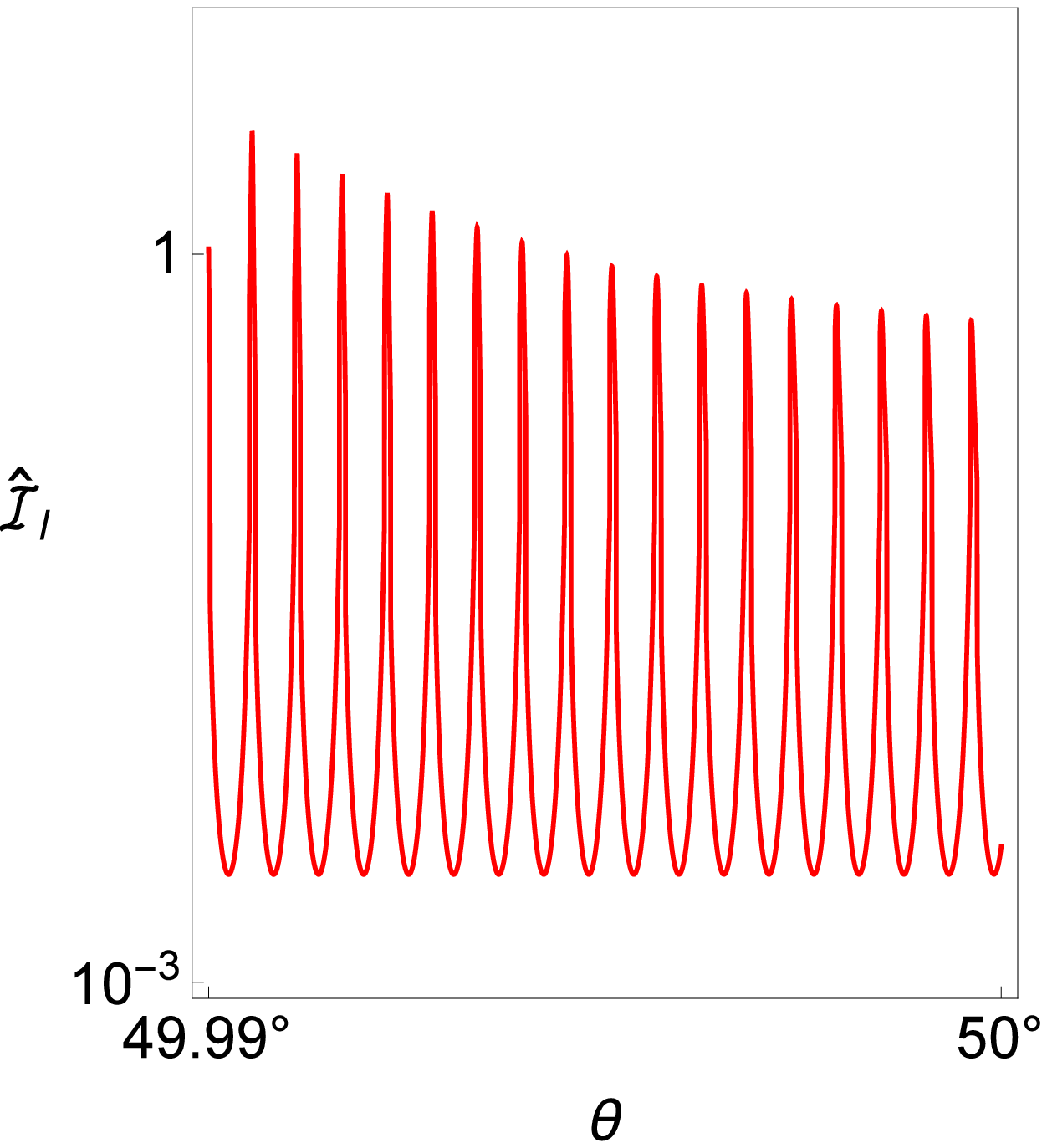}~~~
    \includegraphics[scale=.45]{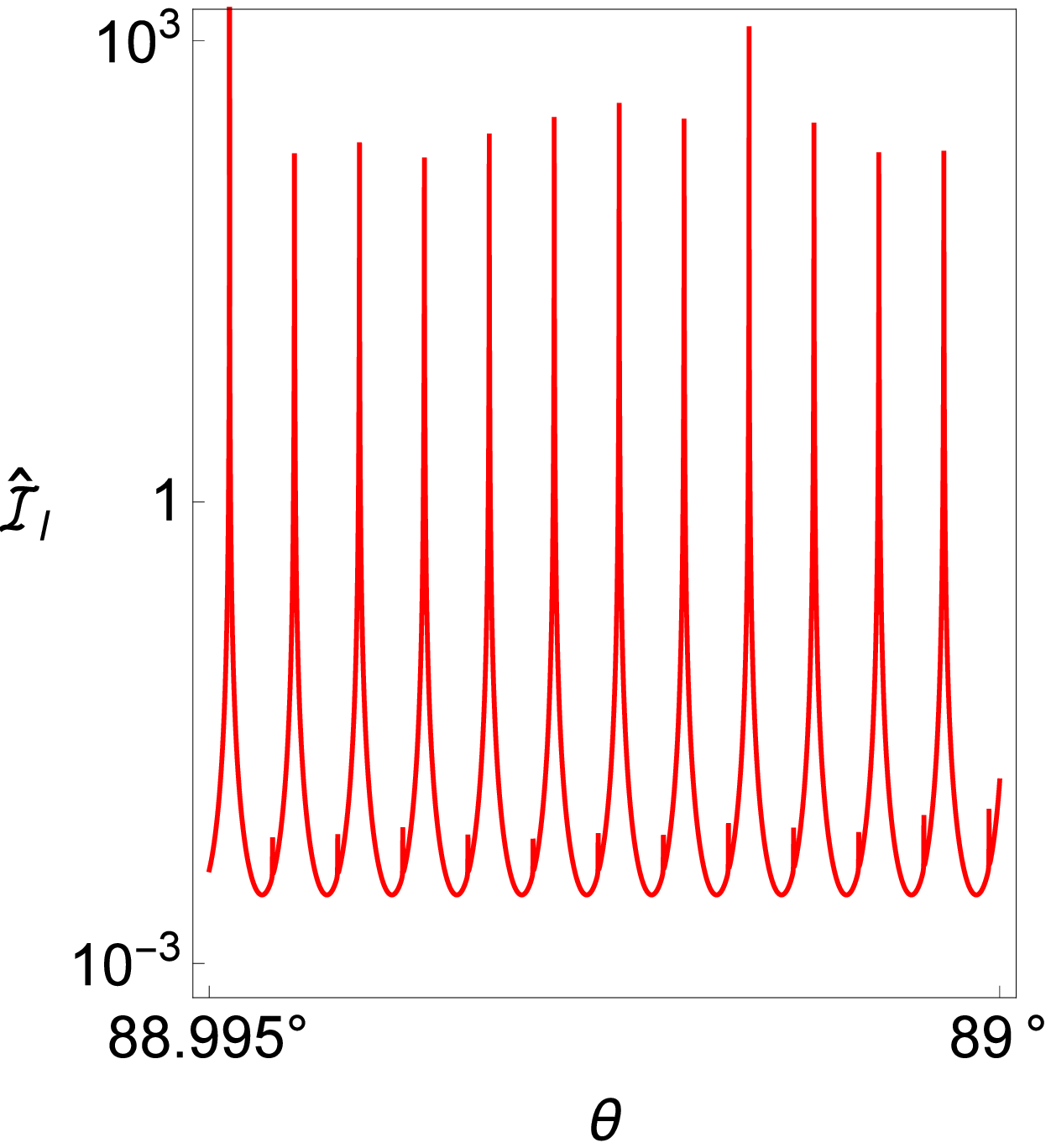}~~~
    \caption{(Color online) Logarithmic plots of $\widehat I_{l}$ as a function of $\theta$. Values of the relevant physical parameters are given by (\ref{specifics}) and (\ref{specifics-mirrors}).}
    \label{fign5}
    \end{center}
    \end{figure}

Fig.~\ref{fign5} describes the dependence of $\widehat I_{l}$ on the incidence angle $\theta$. $\widehat I_{l}$ turns out to oscillate rapidly as one changes $\theta$.

\begin{figure}
    \begin{center}
    \includegraphics[scale=.60]{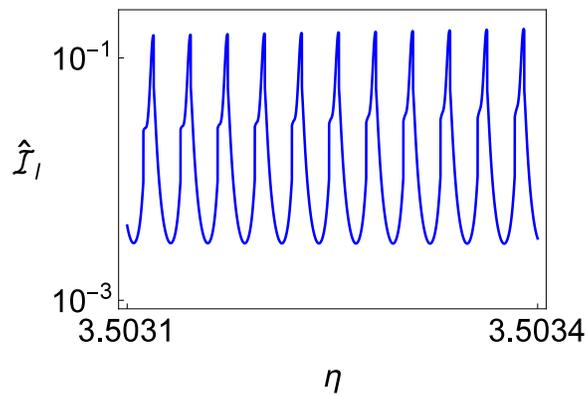}
    \caption{(Color online) Logarithmic plot of $\widehat I_{l}$ as a function of $\eta$ for
    $\theta = 0^{\circ}$. Values of the other relevant physical parameters are given by (\ref{specifics}) and (\ref{specifics-mirrors}).}
    \label{fign3}
    \end{center}
    \end{figure}

Finally, Fig.~\ref{fign3} shows the effect of changing the real part $\eta$ of the refractive index of the gain material on $\widehat I_l$. Again $\widehat I_l$ experiences rapid oscillations as one changes $\eta$.

\section{Discussion and Conclusion}

Textbooks usually offer the following formula for the threshold gain for the normally incident TE modes of the system we consider in this article \cite{milonni-eberly}.
    \be
    g_0 = - \frac{1}{2 L} \ln (r_1 r_2) + \alpha,
    \label{e15}
    \ee
where $r_1:=|R_1^r|^2$ and $r_2:=|R_2^l|^2$ are reflectance of the mirrors, and $\alpha$ is the distributed loss per unit length which is not associated with the mirrors and may usually be ignored. A simple consequence of (\ref{e15}) is that the threshold gain does not depend on the position of the mirrors. This seems to contradict our results which follow the identification of laser light emission with the production of purely outgoing waves, i.e., Postulate 1 of Sec.~\ref{Sec1}. To clarify the situation, let us set $\theta=0$ in (\ref{th-g=1}) -- (\ref{fa0=}). In this case, $\tilde\fn=\fn$, $\fa_0=1$, and we can express (\ref{th-g=1}) in the form (\ref{e15}) provided that we set
    \be
    \alpha=\frac{1}{L}\ln\left|
    \left[\frac{1+\widehat\fn_0 R_1^r}{1+\widehat\fn_0(R_1^r)^{-1}}\right]
    \left[\frac{1+\widehat\fn_0 R_2^l e^{-2i\fK_0}}{1+\widehat\fn_0(R_2^l)^{-1}e^{2i\fK_0}}\right]\right|,
    \ee
and introduce $\widehat\fn_0:=(\fn_0-1)/(\fn_0+1)$. Therefore, our analysis shows that $\alpha$ does depend on the properties of the mirrors particularly when $|\fn_0|$ differs from 1 substantially. For typical gas lasers $|\fn_0|\approx 1$, $|\widehat\fn_0|\ll 1$ and we can ignore $\alpha$. This agrees with the standard textbook description of the threshold gain.

Our investigation of laser output intensity relies on both Postulates 1 and 2 of Sec.~\ref{Sec1}. It provides a derivation of the linear dependence of the output intensity on the gain coefficient, i.e., Eq.~(\ref{eq1}), and gives an explicit formula for the intensity slope $\widehat I$. The only free parameter of this construction is the Kerr coefficient $\sigma$ that in a sense stores the information about the microscopic features of lasing phenomenon. The general behavior of the output intensity is described by the dependence of $\widehat I$ on various macroscopic parameters. We have offered a detailed examination of these for our simple system. Our approach can be slightly extended to the cases that the real part of the refractive index of the slab $\eta$ undergoes changes due to the presence of the nonlinearity. For example, one can use a particular dispersion relation to relate $\eta$ to the $k$ and use the information regarding the change in $k_0$, i.e., $k_0\to k_0+\gamma k_1$ to find the change in the value of $\eta_0$, namely $\gamma\eta_1$. As shown in \cite{p131} for a mirrorless slab laser, this procedure gives $\eta_1=\beta\fK_1$, where $\beta$ is a real parameter carrying the information regarding the dispersion properties of the gain medium. The use of $\eta_1=\beta\fK_1$ together with (\ref{g-g0}) again leads to Eq.~(\ref{eq1}) for the output intensity, but now the intensity slope $\widehat I$ also involves $\beta$.

\subsection*{Acknowledgments} This work has been supported by  the Scientific and Technological Research Council of Turkey (T\"UB\.{I}TAK) in the framework of the project no: 114F357, and by the Turkish Academy of Sciences (T\"UBA).

\section*{Appendix A: $g_0^{(s)}$ and $g_0^{(j)}$ for typical high-gain material}

For a slab made of a typical high-gain material with a thickness much larger than the wavelength of the emitted wave, $|\kappa_0|\ll \eta_0-1\ll k_0L$. In view of this relation and Eqs.~(\ref{th-lambda=1}) -- (\ref{fa0=}),
    \bea
    k_0&\approx&\frac{\pi m}{L\sqrt{\eta_0^2-\sin^2\theta}},
    \label{k-zero=}\nn\\
    g_0^{(s)} &\approx& \frac{2\fa_0}{L}\,\ln\left|\frac{\tilde\eta_0+1}{\tilde\eta_0-1}\right|,
    ~~~~~~~~~~\fa_0\approx\frac{\sqrt{\eta_0^2 -\sin\theta^2}}{\eta_0},\nn\\
    g_0^{(1)} &\approx&\frac{\fa_0}{L}\,
    \ln\left|\frac{(\tilde\eta_0+1)-(\tilde\eta_0-1)R_1^{r}}{(\tilde\eta_0-1)+
    (\tilde\eta_0+1)R_1^{r}}\right|-\frac{g_0^{(s)}}{2},\nn\\
    g_0^{(2)} &\approx&\frac{\fa_0}{L}\ln\left|\frac{\fb_0+4R_2^{l} (1+R_2^{l2})(\tilde\eta_0^{2}+1)\mathfrak{a}_1 + 2R_2^{l2}(\tilde\eta_0^{2}-1)\mathfrak{a}_2}{\sqrt{\mathfrak{a}_3\mathfrak{a}_4}} \right|-\frac{g_0^{(s)}}{2},
    \nn
    \eea
where `$\approx$' labels approximate equalities in which we neglect terms of order $\kappa_0^\ell$ and $(\kappa_0/m)^{\ell-1}$ with $\ell\geq 2$, and
    \bea
    \tilde\eta_0&:=& \sec\theta\sqrt{\eta_0^2 - \sin\theta^2},~~~~
    \tilde\kappa_0:=\frac{\eta_0\kappa_0\sec\theta}{\sqrt{\eta_0^2-\sin\theta^2}},\nn\\
    \fb_0&:=&(1+R_2^{l4})(\tilde\eta_0^{2}-1)^2+ 4R_2^{l2}(\tilde\eta_0^{2}+1)^2,
    \nn\\
    \mathfrak{a}_1 &=& (\tilde\eta_0^{2}-1)\cos\left(\frac{2\pi m}{\tilde\eta_0}\right)+
    \tilde\kappa_0  \sin\left(\frac{2\pi m}{\tilde\eta_0}\right),\nn\\
    \mathfrak{a}_2 &=& (\tilde\eta_0^{2}-1)\cos\left(\frac{4\pi m}{\tilde\eta_0}\right)+2\tilde
    \kappa_0 \sin\left(\frac{4\pi m}{\tilde\eta_0}\right),\notag\\
    \mathfrak{a}_3 &=& (\tilde\eta_0-1)^2 R_2^{l}\left[R_2^{l} + 2\cos\left(\frac{2\pi m}{\tilde
    \eta_0}\right)\right]+(\tilde\eta_0+1)^2,\notag\\
    \mathfrak{a}_4 &=& \left\{(\tilde\eta_0-1)^2\left[1 + 2R_2^{l}\cos\left(\frac{2\pi m}{\tilde      \eta_0}\right)\right]+(\tilde\eta_0+1)^2 R_2^{l2}\right\}^3. \notag
    \eea

\section*{Appendix B: Calculation of $\partial_{\fK_0} G_{+}^{(0)}$, $\partial_{\fn_0} G_{+}^{(0)}$, and $G_{+}^{(1)}$}

According to (\ref{e3}), (\ref{linearsoln2}), and (\ref{gplus}),
    \be
    G^{(0)} (\fn, \fK) = \frac{i A_5 \fK}{2\tilde{\fn}T_2} \left(V'_{+} U'_{+}
    e^{-i \fK\,\tilde{\fn}} - V'_{-} U'_{-} e^{i\fK\,\tilde{\fn}}\right),
    \label{gplus2}
    \ee
where $U'_{\pm}$ and $V'_{\pm}$ are given by (\ref{upm})  and (\ref{vpm}) with $\fn_0$ and $\fK_0$ replaced by $\fn$ and $\fK$, respectively. In view of (\ref{reflections-transmissions}) and
(\ref{gplus2}),
    \begin{align}
    \partial_{\fn_0} G_{+}^{(0)} (\fn_0, \fK_0) &=
    \frac{A_5 \fK_0 \fn_0 V_{-} U_{-}e^{i\fK_0 \tilde{\fn}_0}}{T_2 (\fn_0^2 - \sin\theta^2)}
    \left[\fK_0 - \frac{2i(\mathfrak{b}_{-}\,\tilde{\fn}_0^2 + \mathfrak{b}_{+})}{V_{-}V_{+}U_{-}U_{+}}\right],
    \label{partial-n0-g}\\
    \partial_{\fK_0} G_{+}^{(0)} (\fn_0, \fK_0) &=
    \frac{A_5 \fK_0 V_{-} e^{i\fK_0 \tilde{\fn}_0}}{T_2 U_{+}} \left[(\tilde{\fn}_0^2 - 1) \left(e^{i\fK_0} + R_{2}^{l}\,e^{-i\fK_0}\right)^2 + 4 \mathbf{b}' R_2^{l} + \frac{4 \mathbf{a} R_1^{r} U_{-}U_{+}}{V_{-}V_{+}}\right],
    \label{partial-k0-g}
    \end{align}
where $\mathfrak{b}_{\pm} := R_2^{l} + (R_1^{r} \pm 1)\,e^{2i\fK_0} - R_1^{r}R_2^{l} (R_1^{r} + R_2^{l} \mp R_1^{r} R_2^{l})\,e^{-2i\fK_0}$.

Next, we compute $G_{+}^{(1)} (\fn_0, \fK_0)$. In terms of
     \be
     G_{\pm}^{(1)}(0):= \zeta'_{1}(0) \pm i\fK \zeta_{1}(0),
     \label{gpm1}
     \ee
it takes the form: $G_{+}^{(1)} (\fn_0, \fK_0) = G_{+}^{(1)} (0) + G_{-}^{(1)} (0) R_1^{r}$. We can use (\ref{zeta1}), (\ref{lss}), and (\ref{gpm1}) to evaluate the right-hand side of this relation. This gives
     \be
     G_{+}^{(1)} (\fn_0, \fK_0) = \frac{A_5 \left|A_5\right|^2 V_{-} e^{i\fK_0\tilde{\fn}_0} }{U_{+}}\int_{1}^{0}\,h^2 (\bz') \left|h (\bz')\right|^2 d\bz',
     \label{G+-2=}
     \ee
where $h (\bz) := \left[ U_{+}e^{i\fK_0\tilde{\fn}_0(\bz - 1)} + U_{-}e^{-i\fK_0\tilde{\fn}_0(\bz - 1)} \right] /2\tilde{\fn}_0$. Performing the integral in (\ref{G+-2=}) and making repeated use of (\ref{lss}), we find
     \begin{align}
     G_{+}^{(1)} (\fn_0, \fK_0)&=
     \frac{-iA_5 \left|A_5\right|^2 V_{-} e^{i\fK_0\tilde{\fn}_0}}{16 \fK_0 \tilde{\fn}_0^{3}\tilde{\fn}^{\ast}_0 U_{+}} \Big\{ 3\left[U_{-}^2\,\left|U_{+}\right|^2\,\mathcal{F}_{+} (\tilde{\fn}_0) - U_{+}^2\,\left|U_{-}\right|^2\,\mathcal{F}_{-} (\tilde{\fn}_0)\right]\notag\\ &+ 3U_{-}U_{+}\left[\left|U_{+}\right|^2\,\mathcal{F}_{+} (-\tilde{\fn}_0) - \left|U_{-}\right|^2\,\mathcal{F}_{-} (-\tilde{\fn}_0)\right] +
      U_{+}^{\ast} U_{-}^{3} \mathcal{F}_{+} (3\tilde{\fn}_0) \notag\\ &
      - U_{-}^{\ast} U_{+}^{3} \mathcal{F}_{-} (3\tilde{\fn}_0) + U_{+}^2\,\left|U_{+}\right|^2\,\mathcal{F}_{+} (-3\tilde{\fn}_0) - U_{-}^2\,\left|U_{-}\right|^2\,\mathcal{F}_{-} (-3\tilde{\fn}_0)
     \Big\},
     \label{gplusm1}
     \end{align}
where $\mathcal{F}_{\pm} (\fa) := [1 - e^{\pm i \fK_0 (\tilde{\fn}_0^{\ast} + \fa)}]/(\tilde{\fn}_0^{\ast} + \fa)$.

\ed